\documentclass[journal=jctcce,manuscript=article,layout=twocolumn]{achemso}
\setkeys{acs}{articletitle=true}

\usepackage{chemformula} 
\usepackage[T1]{fontenc} 
\usepackage{amsmath,amsfonts,amsthm}
\usepackage{physics}
\usepackage{comment}
\usepackage{booktabs}
\usepackage{multirow}
\usepackage{textcomp}  
\mciteErrorOnUnknownfalse
\usepackage{subfigure}
\usepackage{enumerate}
\usepackage{colortbl}

\renewcommand{\vec}[1]{\boldsymbol{#1}}

\newcommand{\card}[1]{\abs{ #1 }}
\newcommand\orth[1]{\,\mathrm{orth}\bigl(#1\bigr)}
\newcommand{\diff}{\mathrm{d}}

\newcommand*\bigO{O}

\newcommand*\bbR{\mathbb{R}}
\newcommand*\calD{\mathcal{D}}
\newcommand*\calI{\mathcal{I}}
\newcommand*\calU{\mathcal{U}}

\newtheorem{theorem}{Theorem}[section]
\newtheorem{remark}[theorem]{Remark}

\newcommand*\oriHop{\widehat{H}}
\newcommand*\Hop{\widetilde{H}}

\renewcommand{\op}[1]{\hat{\boldsymbol{#1}}}

\newcommand{\optilde}[1]{\tilde{\boldsymbol{#1}}}

\newcommand*\nelec{n_{\text{e}}}
\newcommand*\morb{M}
\newcommand*\norb{N}
\newcommand*\Uspace{\calU(\morb, \norb)}
\newcommand*\psiset{\{\psi_1, \dots, \psi_\morb\}}
\newcommand*\phiset{\{\phi_1, \dots, \phi_\norb\}}
\newcommand*\Dphi{\calD[(\phi_1, \dots, \phi_\norb)]}
\newcommand*\Dpsi{\calD[(\psi_1, \dots, \psi_\morb)]}
\newcommand*\DpsiU{\calD[(\psi_1, \dots, \psi_\morb)U]}
\newcommand*\oneD{{}^1D}
\newcommand*\twoD{{}^2D}

\definecolor{lightgray}{gray}{0.9}

\author{Yingzhou Li}
\email{yingzhou.li@duke.edu}
\author{Jianfeng Lu}
\email{jianfeng@math.duke.edu}
\affiliation[Duke]{Department of Mathematics, Duke University}
\alsoaffiliation{Department of Chemistry and Department of Physics,
Duke University}

\title[Optimal Orbital Selection FCI] {Optimal Orbital Selection for Full
Configuration Interaction (OptOrbFCI): Pursuing the Basis Set Limit under a
Budget}

\abbreviations{FCI, CDFCI, OptOrbFCI, RDM, 2RDM, CASSCF, BB, HF}

\keywords{orbital selection, basis set limit, full configuration
interaction, CASSCF, ground-state energy; eigenvalue}

\begin{document}

%
%
%
%
%

\begin{abstract}
  Full configuration interaction (FCI) solvers are limited to small basis
  sets due to their expensive computational costs.  An optimal orbital
  selection for FCI (OptOrbFCI) is proposed to boost the power of existing
  FCI solvers to pursue the basis set limit under a computational budget.
  The optimization problem coincides with that of the complete active space
  SCF method (CASSCF), while OptOrbFCI is algorithmically quite different.
  OptOrbFCI effectively finds an optimal rotation matrix via solving a
  constrained optimization problem directly to compress the orbitals of
  large basis sets to one with a manageable size, conducts FCI calculations
  only on rotated orbital sets, and produces a variational ground-state
  energy and its wave function.  Coupled with coordinate descent full
  configuration interaction (CDFCI), we demonstrate the efficiency and
  accuracy of the method on the carbon dimer and nitrogen dimer under basis
  sets up to cc-pV5Z. We also benchmark the binding curve of the nitrogen
  dimer under the cc-pVQZ basis set with 28 selected orbitals, which provide
  consistently lower ground-state energies than the FCI results under the
  cc-pVDZ basis set. The dissociation energy in this case is found to be of
  higher accuracy.
\end{abstract}

\section{Introduction}
\label{sec:introduction}

Quantum many-body problems in electronic structure calculations remain
difficult for strongly correlated (multireference) systems. Both
the infamous sign problem and the combinatorial scaling make the
problem intractable in a large basis set setting.  In this paper, we
propose an optimal orbital selection for FCI (OptOrbFCI) to solve
full configuration interaction (FCI) problems on large basis sets
under limited memory and computational power budget.

In the past decades, methods for solving FCI problems have been developed
rapidly, which gives an acceleration of a factor of hundreds or even more
compared with conventional methods. Among these efficient FCI solvers, the
density matrix renormalization group (DMRG)~\cite{Chan2011,
Olivares-Amaya2015} employs a matrix product state ansatz in representing
the ground-state wave function and then finds variational solutions. Full
configuration interaction quantum Monte Carlo (FCIQMC)~\cite{Booth2009,
Booth2012} and its variants (iFCIQMC~\cite{Cleland2010},
S-FCIQMC~\cite{Petruzielo2012}) adopt the stochastic walker representation
of wave functions in the second quantization which is updated in each
iteration according to the Hamiltonian operator; convergence is guaranteed
in the sense of inexact power method~\cite{Lu2020}. Configuration
interaction by perturbatively selecting iteration (CIPSI)~\cite{Huron1973},
adaptive configuration interaction (ACI)~\cite{Schriber2016}, adaptive
sampling configuration interaction (ASCI)~\cite{Tubman2016, Tubman2018a},
heat-bath configuration interaction (HCI)~\cite{Holmes2016}, and stochastic
HCI (SHCI)~\cite{Sharma2017} dynamically select important configurations
according to various approximations of the perturbation and then provide
variational solutions via traditional eigensolvers together with a post
perturbation estimation of the ground-state energy. Coordinate descent full
configuration interaction (CDFCI)~\cite{Wang2019} reformulates the FCI
problem as an unconstrained optimization problem and variationally solves it
via coordinate descent method with hard thresholding. The systematic full
configuration interaction fast randomized iteration
(sFCI-FRI)~\cite{Greene2019} applies a fast randomized iteration
framework~\cite{Lim2017} to FCI problems and introduces a hierarchical
factorization to further reduce the computational cost. Several other
methods~\cite{Lim2017, Li2019c, Hernandez2019, Gao2020} attempting to solve
FCI problems are developed from the numerical linear algebra community.
Nevertheless, none of the aforementioned methods can give accurate results
for basis sets of size beyond a few dozen, due to the exponential scaling of
the computational cost with respect to the basis set size.

FCI solvers, viewed as post-Hartree-Fock (HF) methods, usually
adopt molecular orbitals (one-electron and two-electron integrals)
from HF calculation and solve the many-body problem starting from
there. Thanks to the rotation applied to the basis set (in most
cases atomic orbitals) in HF calculation, the molecular orbitals
usually give compressible representation of the many-body wave
function. In order to further boost the compressibility, one
may consider embedding the FCI solver in another loop of orbital
rotation.~\cite{Tubman2018a} The procedure used in~\citet{Tubman2018a}
can be described as follows. Given a set of orbitals, they first apply
the FCI solver to generate a rough approximation of the ground-state
wave function and its associated one-body density matrix (1RDM). Then
these orbitals are rotated via the eigenvectors of the 1RDM. The
rotated orbitals are known as the natural orbitals. Using the rotated
orbitals~(rotated one-body and two-body integrals), the FCI solver
is applied again. These two steps are performed repeatedly until
some stopping criterion is achieved. This procedure aims to produce
orbitals with better compressibility in representing the many-body
wave function. The optimality of the natural orbital has been questioned
in several works~\cite{Bytautas2003, Zhang2014, Giesbertz2014,
Alcoba2016}, which proposed various optimization procedures under
different definitions of optimalities.  One shortcoming of all
these works, however, is that all these orbital rotations build on
top of the many-body wave function with orbitals of the same size as
that of the original molecular orbitals; thus it does not save much
computational cost when we start with a large basis set.

In this paper, we consider the following problem: Given a large basis set
and limited memory and computational power, what is the optimal variational
ground-state energy under the FCI framework? More specifically, let us
consider a system with $\nelec$ electrons. An HF calculation with a basis
set provides the molecular orbitals of size $\morb$, $\psiset$.  Under the
restriction of memory usage and computational power, we assume that the FCI
solver is only able to solve the FCI problem with $\norb$ orbitals, where
$\norb < \morb$. Our goal is then to find a partial unitary matrix $U \in
\bbR^{\morb \times \norb}$ such that the ground-state energy is minimized
under an optimal set of orbitals of size $\norb$, generated from the partial
unitary transform of $\psiset$ via $U$. For simplicity we assume that the
orbitals are real valued functions and the partial unitary matrix is a real
matrix.  Such an optimal orbital selection procedure is not only valuable to
FCI computations on classical computers but also to FCI computations on
noisy intermediate-scale quantum computers.~\cite{Kivlichan2018,
Babbush2019} Due to the limited number of computational qubits in current
quantum computers, compression of orbitals is very much desired.

Although starting from different perspectives, this problem ends up pursuing
the same goal as the complete active space self-consistent field
method~(CASSCF)~\cite{Siegbahn1980, Roos1980, Siegbahn1981,
Knowles1985,Zgid2008, Ghosh2008, Yanai2009, Olsen2011, Wouters2014,
LiManni2016, Leon2017, Ma2017, Smith2017, Sun2017, Freitag2019, Kreplin2019,
Levine2020}. CASSCF is a complete active space version of
multiconfigurational self-consistent field (MCSCF) method, which aims to
extend the Hartree-Fock calculation to multi-configurational spaces. Hence,
comparing CASSCF and the goal of this paper, CASSCF is proposed starting
from extending the Hartree-Fock computational whereas the latter is proposed
starting from compressing the FCI computation. Both reach the same place.
CASSCF has been rapidly developed for several decades. There are two popular
algorithms~\cite{Siegbahn1980, Olsen2011}, \latin{i.e.}, the super-CI
method~\cite{Ruedenberg1979, Roos1980} and the Newton
method~\cite{Siegbahn1981}. The super-CI method solves the first order
variational condition with respect to the FCI coefficients and
orbitals~\cite{Ruedenberg1979}, and results in solving an FCI problem in the
active space and an eigenvalue problem in parametrized singly excited
states.  The Newton method converts the problem to an unconstrained
optimization problem and solves it using the Newton method. Since both
methods adopt local approximation of the rotation matrix, efficiency is
guaranteed only locally.  (The two methods will be recalled and presented
from an optimization point of view below.) Recently, several modified
schemes are developed to further accelerate the orbital
minimization~\cite{Yanai2009, Sun2017, Kreplin2019}.  Other related
developments in CASSCF replace the direct FCI solver by the modern FCI
solvers mentioned above~\cite{Zgid2008, Ghosh2008, Yanai2009, Wouters2014,
Leon2017, Ma2017, Freitag2019, LiManni2016, Smith2017, Levine2020}. Although
targeting the same problem as CASSCF, since the starting points are quite
different, we pursue effective algorithms under the setting that FCI solvers
are computationally much more expensive compared to the orbital
optimization.  Such a setting is natural when applying modern FCI solvers to
large active orbital spaces and when solving FCI problems on a quantum
computer.  Our proposed formulas and algorithm, hence, are different from
conventional CASSCF algorithms~\cite{Roos1980, Siegbahn1980}. Instead of
proposing an ansatz for the rotation matrix and truncating the expression,
we optimize the rotation matrix directly through a constrained optimization
solver such that the orbital optimization can converge to a minimizer far
away from the initial point achieving a better energy.  The better orbital
optimization potentially reduces the number of macro iterations, which is
the total number of solving FCI problems in the active space, and avoids
some local minima. Numerically, we find that the macro iteration number in
our method either is reduced or remains unchanged comparing to that of
CASSCF. In our experiments, ground-state energies obtained by OptOrbFCI are
always equal or lower than those of CASSCF.

The contribution of our work can be summarized into three parts. First, we
mathematically formulate the problem as a constrained optimization problem
with two variables: a partial unitary matrix $U$ and the ground-state wave
function. Since these two variables are coupled together, the optimization
problem is very difficult to solve directly. Hence we adopt the alternating
minimization idea. The optimization problem is then decoupled into two
single variable optimization problems and solved in an alternating way.
Second, we propose an efficient algorithm, namely OptOrbFCI, for the
optimization problem based on the trials of several possible solvers for
each of the single variable optimization problems. Specifically
CDFCI~\cite{Wang2019} is applied as the FCI solver, which has not been
applied in CASSCF before. Finally, we apply the algorithm to the water
molecule, carbon dimer, and nitrogen dimer. Limited by the size of the
cc-pVDZ basis set,~\footnote{The number of molecular orbitals from the HF
calculation with cc-pVDZ basis set.} we produce the variational ground-state
energy using the optimal orbitals selected from the cc-pVTZ, cc-pVQZ, and
cc-pV5Z basis sets. In all cases, significant improvements of accuracy have
been observed. Moreover, the binding curve of the nitrogen dimer is produced
using the optimal orbitals selected from the cc-pVQZ basis set limited to
the size $28$.  The dissociation energy is much more accurate than the FCI
results under the cc-pVDZ basis set.

The rest of the paper is organized as follows. Section~\ref{sec:formulation}
formulates the constrained optimization problem together with two single
variable subproblems. The detailed algorithm is introduced in
Section~\ref{sec:algorithm}. In Section~\ref{sec:numerical results}, we
apply OptOrbFCI to the water molecule, carbon dimer, and nitrogen dimer to
demonstrate the efficiency of the algorithm. Finally,
Section~\ref{sec:conclusion} concludes the paper together with a discussion
of future work.

\section{Formulation}
\label{sec:formulation}

This section formulates the problem raised in the Introduction as an
optimization problem and derives the related two subproblems.

We first introduce notations used throughout this paper. As before,
$\morb$ and $\norb$ denote the number of the given molecular orbitals
and the computationally affordable number of orbitals ($\norb <
\morb$). The given large orbital set is $\psiset$ and the associated
Hamiltonian operator in the second quantization is
\begin{equation} \label{eq:oriHam}
    \oriHop = \sum_{p,q = 1}^\morb h_{pq} \op{c}_p^\dagger \op{c}_q +
    \frac{1}{2} \sum_{p,q,r,s = 1}^\morb v_{pqrs} \op{c}_p^\dagger
    \op{c}_q^\dagger \op{c}_s \op{c}_r,
\end{equation}
where $\op{c}_p^\dagger$ and $\op{c}_q$ are the creation and
annihilation operators associated with $\psi_p$ and $\psi_q$
respectively.  The one-electron and two-electron integrals, $h_{pq}$
and $v_{pqrs}$, admit the following expressions,
\begin{align}
    h_{pq} = & \int \diff \vec{x}_1 \, \psi^\star_p(\vec{x}_1) h(\vec{x}_1)
    \psi_q(\vec{x}_1), \text{ and}\\
    \begin{split}
        v_{pqrs} = & \int \diff \vec{x}_1 \diff \vec{x}_2 \,
        \psi^\star_p(\vec{x}_1) \psi^\star_q(\vec{x}_2) \\
        & \hspace{3em} \cdot v(\vec{x}_1,
        \vec{x}_2) \psi_s(\vec{x}_2) \psi_r(\vec{x}_1),
    \end{split}
\end{align}
where $h(\vec{x}_1)$ and $v(\vec{x}_1, \vec{x}_2)$ are the one-body and
two-body operators, respectively.  However, due to the limited memory and
computational power, we are only able to solve FCI problems under $\norb$
orbitals. Hence, we introduce a partial unitary matrix $U \in \Uspace$,
where $\Uspace$ is the space of all partial unitary matrix of size $\morb$
by $\norb$, \latin{i.e.},
\begin{equation}
    \Uspace = \big\{U \in \bbR^{\morb \times \norb} \mid U^\top U =
    I_\norb \big\}
\end{equation}
and $I_\norb$ denotes the identity matrix of size $\norb$ by
$\norb$. The transformed orbitals from $\psiset$ via $U$ are denoted
as $\phiset$ such that
\begin{equation} \label{eq:phipsiU}
    \phi_i = \sum_{j=1}^\morb \psi_j U_{ji},
\end{equation}
where $U_{ji}$ denotes the $(j,i)$-th entry of $U$.  We also adopt the
expression $(\phi_1, \dots, \phi_\norb) = (\psi_1, \dots, \psi_\morb)
U$ to denote the transformation.  The Hamiltonian operator associated
with $\phiset$ is then,
\begin{equation} \label{eq:Ham}
    \begin{split}
        \Hop = & \sum_{p',q'= 1}^\norb \tilde{h}_{p'q'}
        \op{d}_{p'}^\dagger \op{d}_{q'} \\
        & + \frac{1}{2} \sum_{p',q',r',s' = 1}^\norb \tilde{v}_{p'q'r's'}
        \op{d}_{p'}^\dagger
        \op{d}_{q'}^\dagger \op{d}_{s'} \op{d}_{r'},
    \end{split}
\end{equation}
where $\op{d}_{p'}^\dagger$ and $\op{d}_{q'}$ are the creation and
annihilation operators associated with $\phi_{p'}$ and $\phi_{q'}$
respectively, the one-electron integral $\tilde{h}_{p'q'}$ is
\begin{equation} \label{eq:hpq}
    \begin{split}
        \tilde{h}_{p'q'} = & \int \diff \vec{x}_1 \,
        \phi^\star_{p'}(\vec{x}_1) h(\vec{x}_1)
        \phi_{q'}(\vec{x}_1)\\
        = & \sum_{p,q = 1}^\morb h_{pq}  U_{pp'} U_{qq'},
    \end{split}
\end{equation}
and the two-electron integral $\tilde{v}_{p'q'r's'}$ is
\begin{equation} \label{eq:vpqrs}
    \begin{split}
        \tilde{v}_{p'q'r's'} = & \int \diff \vec{x}_1 \diff \vec{x}_2 \,
        \phi^\star_{p'}(\vec{x}_1) \phi^\star_{q'}(\vec{x}_2) \\
        & \hspace{3em} \cdot v(\vec{x}_1,
        \vec{x}_2) \phi_{s'}(\vec{x}_2) \phi_{r'}(\vec{x}_1) \\
        = & \sum_{p,q,r,s = 1}^\morb 
        v_{pqrs} U_{pp'} U_{qq'} U_{ss'} U_{rr'}.
    \end{split}
\end{equation}
The connection \eqref{eq:phipsiU} between orbital set $\psiset$ and
$\phiset$ implies the connection between annihilation
operators,
\begin{equation} \label{eq:dcU}
    \op{d}_{q'} = \sum_{q = 1}^{\morb} \op{c}_q U_{qq'}.
\end{equation}
Such a relationship also holds for creation operators.

Moreover, we denote the variational space for wave function as
$\Dphi = \DpsiU$, which is the span of all Slater determinants
constructed from $\phiset$.

With all notations defined above, our problem can be formulated as,
\begin{equation} \label{eq:oriopt-oriH}
    \min_{ \substack{ \ket{\Phi} \in \DpsiU \\ \braket{\Phi} = 1 \\
    U \in \Uspace} } \ev**{\oriHop}{\Phi}.
\end{equation}
Notice the second quantization form of $\oriHop$ is under orbital set
$\psiset$ whereas the wave function $\ket{\Phi}$ lives in the
variational space associated with $\phiset$. Such an inconsistency is
inconvenient to handle numerically.

We now show that it is in fact equivalent to replace the Hamiltonian
$\oriHop$ in \eqref{eq:oriopt-oriH} by $\Hop$; thus, both the Hamiltonian
and the wave function are associated with the same set of orbitals
$\phiset$. The connection between $\op{d}_{q'}$ and $\op{c}_q$
in \eqref{eq:dcU} leads to the anticommutation relation between
$\op{d}_{p'}^\dagger$ and $\op{c}_q$,
\begin{equation} \label{eq:anticommcd}
    \acomm{\op{c}_{q}}{\op{d}_{p'}^\dagger} = \sum_{p =
    1}^{\morb} \acomm{\op{c}_{q}}{\op{c}_{p}^\dagger} U_{pp'} = U_{qp'}.
\end{equation}
Define another operator $\optilde{c}_{q} = \sum_{q' = 1}^{\norb}
\op{d}_{q'} U_{qq'}$. The anticommutation relation between
$\op{d}_{p'}^\dagger$ and $\optilde{c}_q$ is the same as
\eqref{eq:anticommcd},
\begin{equation} \label{eq:anticommctilded}
    \acomm{\optilde{c}_{q}}{\op{d}_{p'}^\dagger} = \sum_{q' =
    1}^{\norb} \acomm{\op{d}_{q'}}{\op{d}_{p'}^\dagger} U_{qq'}
    = U_{qp'}.
\end{equation}
Since both $\op{c}_q$ and $\optilde{c}_q$ have the same
anticommutation relation with $\op{d}_{p'}^\dagger$, these two
annihilation operators acting on any wave function $\ket{\Phi}$
in $\DpsiU$ give the same results, \latin{i.e.},
\begin{equation} \label{eq:cctildePhi}
    \op{c}_q \ket{\Phi} = \optilde{c}_q \ket{\Phi}.
\end{equation}
Hence, the objective function $\ev**{\oriHop}{\Phi}$ in
\eqref{eq:oriopt-oriH} admits the same result if all creation and
annihilation operators are replaced by $\optilde{c}_p^\dagger$ and
$\optilde{c}_q$. The resulting Hamiltonian is exactly $\Hop$
associated with $\phiset$ defined in \eqref{eq:Ham}. A more detailed 
derivation can be found in Appendix~\ref{app:equivalentHam}. Our
problem \eqref{eq:oriopt-oriH}, thus, is equivalent to,
\begin{equation} \label{eq:oriopt}
    \min_{ \substack{ \ket{\Phi} \in \DpsiU \\ \braket{\Phi} = 1 \\
    U \in \Uspace} } \ev**{\Hop[U]}{\Phi},
\end{equation}
where $\Hop[U]$ is $\Hop$ defined in \eqref{eq:Ham} and we write
$U$ in brackets to emphasize its dependency on $U$.

\begin{remark} \label{rmk:obj}
    If we assume that under optimal orbital selection, the system with
    a smaller number of electrons has higher energy, then it can be shown
    that \eqref{eq:oriopt} is equivalent to the following problem:
    \begin{equation} \label{eq:genopt}
      \min_{ \substack{ \ket{\Psi} \in \Dpsi \\ \braket{\Psi} = 1 \\
          U \in \Uspace} } \ev**{\Hop[U]}{\Psi},
    \end{equation}
    where the wave function $\ket{\Psi}$ now lives in a larger
    variational space (and thus the computational cost exceeds
    the limitation). We shall focus on the surrogate problem
    \eqref{eq:oriopt}, which is computationally feasible.
\end{remark}

The objective function in our original problem \eqref{eq:oriopt-oriH} has
the same expression as that in the FCI problem under the orbital set
$\psiset$.  Moreover, any feasible wave function in \eqref{eq:oriopt-oriH}
belongs to the space $\Dpsi$, which is the variational space of the FCI
problem under $\psiset$. Since FCI problem under $\psiset$ is a variational
method for the many-body Schr{\"o}dinger equation, our problem
\eqref{eq:oriopt-oriH} is also a variational method and so is
\eqref{eq:oriopt}. Therefore, solving \eqref{eq:oriopt} gives a variational
ground-state energy and its wave function.

We see that $\ket{\Phi}$ and $U$ in \eqref{eq:oriopt} are coupled together.
Instead of minimizing $\ket{\Phi}$ and $U$ simultaneously, we minimize
\eqref{eq:oriopt} in an alternating fashion. We first fix $U$ and minimize
\eqref{eq:oriopt} with respect to $\ket{\Phi}$ only. Once the minimizer of
$\ket{\Phi}$ is achieved, we then fix $\ket{\Phi}$ and minimize
\eqref{eq:oriopt} with respect to $U$ only. The procedure is repeated until
some convergence criterion is achieved.  Next, we derive the two subproblems
for fixed $U$ and fixed $\ket{\Phi}$ respectively.

\subsection*{Subproblem with Fixed $U$.}

When we fix $U$ in \eqref{eq:oriopt}, the orbital set $\phiset$
is also fixed.  The optimization problem \eqref{eq:oriopt} is then
simplified as,
\begin{equation} \label{eq:subopt1}
    \min_{ \substack{ \ket{\Phi} \in \Dphi \\ \braket{\Phi} = 1} }
    \ev**{\Hop}{\Phi},
\end{equation}
which is a standard FCI problem under the orbital set $\phiset$.

\subsection*{Subproblem with Fixed $\ket{\Phi}$.}

When we fix $\ket{\Phi}$, the objective function in \eqref{eq:oriopt}
can be written as,
\begin{equation} \label{eq:subobj2}
    \begin{split}
        \ev**{\Hop[U]}{\Phi} = & \sum_{p',q' = 1}^\norb 
        \tilde{h}_{p'q'} \ev**{\op{d}_{p'}^\dagger \op{d}_{q'}}{\Phi} \\
        & \hspace{-6em} + \sum_{p',q',r',s' = 1}^\norb
        \tilde{v}_{p'q'r's'} \ev**{\op{d}_{p'}^\dagger
        \op{d}_{q'}^\dagger \op{d}_{s'} \op{d}_{r'}}{\Phi} \\
        = & \sum_{p',q' = 1}^\norb \sum_{p,q = 1}^\morb h_{pq} U_{pp'}
        U_{qq'} \oneD^{p'}_{q'} \\
        & \hspace{-8em} + \sum_{p',q',r',s' = 1}^\norb \sum_{p,q,r,s
        = 1}^\morb v_{pqrs} U_{pp'} U_{qq'} U_{rr'} U_{ss'}
        \twoD^{p'q'}_{r's'} \\
        =: & P_4(U),
    \end{split}
\end{equation}
where $\oneD^{p'}_{q'} = \ev**{\op{d}_{p'}^\dagger \op{d}_{q'}}{\Phi}$
and $\twoD^{p'q'}_{r's'} = \ev**{\op{d}_{p'}^\dagger
\op{d}_{q'}^\dagger \op{d}_{s'} \op{d}_{r'}}{\Phi}$ are the standard
one-body reduced density matrix (1RDM) and two-body reduced density
matrix (2RDM) respectively. The objective function, denoted as
$P_4(U)$, is then a fourth order polynomial of $U$. Notice that
$h_{pq}$ and $v_{pqrs}$ are given coefficients associated with the
original molecular orbital set $\psiset$, and $\oneD^{p'}_{q'}$ and
$\twoD^{p'q'}_{r's'}$ are also independent of $U$ as long as we fix
$\ket{\Phi}$. Hence the subproblem can be summarized as
\begin{equation} \label{eq:subopt2}
    \min_{ U \in \Uspace } P_4(U),
\end{equation}
which minimizes a fourth order polynomial of $U$ with an orthonormality
constraint.

\section{Algorithm}
\label{sec:algorithm}

In this section, we will first discuss algorithms for
solving \eqref{eq:subopt1} and \eqref{eq:subopt2} in
Section~\ref{sec:fcisolvers} and Section~\ref{sec:polysolvers}
respectively. Then the overall algorithm, OptOrbFCI, is summarized
as a pseudo code in Section~\ref{sec:overallalgorithm} together with
some discussion on initial guesses, convergence, stopping criteria,
and computational complexities.

\subsection{FCI Solvers and RDM Methods}
\label{sec:fcisolvers}

Algorithms in this section aim for solving the FCI problem
\eqref{eq:subopt1} and producing 1RDM and 2RDM as inputs for
\eqref{eq:subopt2}. Most FCI solvers can produce RDMs.  The potential
choices then include but not limited to, DMRG~\cite{Chan2011,
Olivares-Amaya2015}, FCIQMC~\cite{Booth2009}, ACI~\cite{Schriber2016},
HCI~\cite{Holmes2016}, and CDFCI~\cite{Wang2019}. The perturbation energy is
not needed for intermediate iterations and is optional for the last FCI
solved in OptOrbFCI.  Throughout this paper, CDFCI is the solver used to
address all FCI problems.

Regarding 1RDM and 2RDM, the computational cost is on the same order as
applying the Hamiltonian operator to the many-body wave function one time
while, due to the efficiency of CDFCI, the runtime for the FCI solving part
is also of the same order. Hence the computation of RDMs needs to be
carefully addressed. Since 1RDM can be easily reduced from 2RDM with cheap
computational cost, we focus only on the computation of 2RDM here. Assume
the wave function is of the form $\ket{\Phi} = \sum_{i \in \calI}x_i
\ket{D_i}$, where $\ket{D_i}$ denotes a Slater determinant in $\Dphi$, $x_i$
is the corresponding coefficient, and $\calI$ denotes the index set of
nonzero coefficients, \latin{i.e.}, $x_i \neq 0$ for all $i \in \calI$. We
introduce two methods for computing 2RDM.

The first method is of quadratic scaling with respect to the cardinality of
$\calI$, $\card{\calI}$. It loops over all pairs of Slater determinants with
nonzero coefficients, \latin{i.e.}, $\big(\ket{D_i}, \ket{D_j}\big)$ for
$i,j \in \calI$. If two Slater determinants differ by more than two
orbitals, then this pair does not contribute to 2RDM.  Otherwise, the
contribution to 2RDM is evaluated. Notice that there are only $\bigO(\norb^2
\card{\calI})$ pairs that contribute to 2RDM and all of the rest of the
pairs only require an ``XOR'' and a ``POPCOUNT''~\footnote{Population count
operation counts the number of set bits in a value, which is usually
implemented using hardware in modern computers.} operation, both of which
are of great efficiency in modern computers.

The second method is of linear scaling with respect to $\card{\calI}$. It
loops over all Slater determinants with nonzero coefficients. For each
determinant, $\ket{D_i}$, it applies all possible $\op{d}_{p'}^\dagger
\op{d}_{q'}^\dagger \op{d}_{s'} \op{d}_{r'}$ to the determinant and queries
the coefficient of $\op{d}_{p'}^\dagger \op{d}_{q'}^\dagger \op{d}_{s'}
\op{d}_{r'} \ket{D_i}$. The contribution, \latin{i.e.,} the product of the
coefficients of both determinants and multiplying the sign, is then added to
2RDM. Unlike the first method, where only $\bigO(\card{\calI})$ queries of
the coefficients of the many-body wave function are needed and then these
coefficients are stored and accessed in an array, the second method requires
$\bigO(\norb^2 \card{\calI})$ queries.  In almost all FCI solvers, special
data structures are used to store the wave function with sparse
coefficients, \latin{e.g.}, hash table, black-red tree, sorted array,
\latin{etc.} Querying any of these special data structures is relatively
expensive. Hence the runtime of the second method is much slower than that
of the first one if $\card{\calI}$ is not large.

In practice, we dynamically select the method to compute 2RDM based
on both $\card{\calI}$ and the querying cost. Nevertheless, the
runtime of the second method is guaranteed to be of the same order
as the FCI solving part in CDFCI. Hence the overall total runtime
for solving \eqref{eq:subopt1} and producing RDMs is, in general,
no more than twice of the FCI solver runtime in CDFCI.

\subsection{Optimizing Orthonormal Constrained Polynomial}
\label{sec:polysolvers}

This section introduces the algorithm used to solve \eqref{eq:subopt2}.
Although the objective function is simply a fourth order polynomial of $U$,
the orthonormality constraint makes the problem in general more difficult to
solve than the linear eigenvalue problem.  Luckily, the variable $U$ is only
of dimension $\morb \times \norb$.  Comparing to the FCI problem, which
usually costs $\bigO({\norb \choose \nelec})$ operations, the computational
cost of minimizing \eqref{eq:subopt2}, in most cases, is negligible while
the efficient algorithm is still desired especially when the given molecular
orbital set size $\morb$ is much larger than $\norb$.

Regarding the orthonormality constrained optimization problems, there are
three major groups of techniques to deal with the constraint, namely,
augmented Lagrangian methods~\cite{Wen2013, Gao2019}, projection
methods~\cite{Gao2018}, and manifold based methods~\cite{Zhang2014a,
Huang2015}.  For these methods, we explored the efficiency on a small test
problem and employ a projection method with alternating Barzilai-Borwein
(BB) stepsize~\cite{Gao2018}.

The iteration for the employed method can be written as,
\begin{equation} \label{eq:projBB}
    U_{k+1} = \orth{U_k - \tau_k \grad_U P_4(U_k)},
\end{equation}
where $U_k$ denotes the $U$ matrix at the $k$-th iteration, $\orth{\cdot}$
denotes the orthonormalization function, and $\tau_k$ is the alternating BB
stepsize. The orthonormalization function of any matrix $V$ is defined as
the orthonormal basis of $V$ and implemented as,
\begin{equation*}
    \orth{V} = V Q \Lambda^{-\frac{1}{2}}.
\end{equation*}
where $Q$ and $\Lambda$ are eigenvectors and eigenvalues of $V^\top V$,
\latin{i.e.}, $V^\top V = Q \Lambda Q^\top$. The alternating BB stepsize
applies two BB stepsizes in an alternating way as,
\begin{equation*}
    \tau_k =
    \begin{cases}
        \tau^{\text{BB1}}_k \text{ for odd $k$}\\
        \tau^{\text{BB2}}_k \text{ for even $k$}
    \end{cases},
\end{equation*}
where
\begin{equation*}
    \tau^{\text{BB1}}_k =
    \frac{\langle U_k - U_{k-1}, U_k - U_{k-1} \rangle}{ \abs{\langle
    U_k - U_{k-1}, G_k - G_{k-1} \rangle} },
\end{equation*}
\begin{equation*}
    \tau^{\text{BB2}}_k =
    \frac{\abs{\langle U_k - U_{k-1}, G_k - G_{k-1} \rangle}}{\langle
    G_k - G_{k-1}, G_k - G_{k-1} \rangle},
\end{equation*}
$G_k = \grad_U P_4(U_k)$ is the gradient of $P_4$ at $U_k$, and $\langle A,
B \rangle = \trace{A^\top B}$.

\subsection{OptOrbFCI}
\label{sec:overallalgorithm}

The overall algorithm, OptOrbFCI, hence alternatively minimizes
\eqref{eq:subopt1} and \eqref{eq:subopt2}, with some computations to prepare
the inputs for each other. We summarize OptOrbFCI as follows.

\begin{enumerate}[Step 1]
    \item Set iteration index $k = 0$ and prepare initial guess $U_0$.

    \item Calculate the reduced one-body and two-body integrals
        using $U_k$ as \eqref{eq:hpq} and \eqref{eq:vpqrs}
        respectively.
        \label{alg:firststep}

    \item Solve the FCI problem \eqref{eq:subopt1} via CDFCI method
        and obtain the ground-state wave function and energy.

    \item If the decay of the ground-state energy is smaller than
        the given tolerance, convergence has been achieved and the
        algorithm is stopped.

    \item Compute the 1RDM and 2RDM from the ground-state wave function.

    \item Solve the orthonormal constrained polynomial
        \eqref{eq:subopt2} via projection method with alternating BB
        stepsize as \eqref{eq:projBB} and obtain $U_{k+1}$.

      \item Set $k=k+1$ and repeat Steps
        \ref{alg:firststep}--\ref{alg:laststep}.
        \label{alg:laststep}
\end{enumerate}

Notice in the above algorithm that the stopping criteria are checked right
after the FCI calculation rather than at the end of each iteration. However,
it is not activated until the second iteration so that we can compare the
FCI ground-state energies of the current iteration against those of the
previous iteration. We also emphasize that the CDFCI method employed here is
just one choice of FCI solvers. OptOrbFCI can employ many other FCI solvers
as a replacement.

In the following, we discuss some details of the algorithm,
\latin{i.e.}, initial guesses, convergence, stopping criteria, and
computational complexities.

\subsubsection*{Initial Guesses}

In OptOrbFCI, the only variable needed to be initialized is $U_0$. We found
that using a random orthonormal matrix as the initialization of $U_0$ works
in practice, while, in this case, the FCI ground-state energy in the first
iteration is even worse than the HF energy. A better initialization for
$U_0$, which is the one used throughout all numerical experiments in this
paper, is the permutation matrix selecting $\norb$ different orbitals with
the lowest HF orbital energy from $\psiset$.

Besides the initialization for the overall algorithm, we also need
to give initializations for both subproblems, \eqref{eq:subopt1}
and \eqref{eq:subopt2}. For \eqref{eq:subopt1}, in regular CDFCI, the
wave function is usually initialized as the single HF state. However,
after rotation via $U$, we lose track of the HF state in the new
orbital set, $\phiset$.  Hence we initialize CDFCI as a single state
with $\frac{\nelec}{2}$ orbitals with smallest ``orbital energy''
doubly occupied (spin-up and spin-down), where the ``orbital energy''
of $\phi_i$ is defined as,
\begin{equation}
    \sum_{j}  \varepsilon_j U_{ji}^2,
\end{equation}
where $\varepsilon_j$ is the orbital energy of $\psi_j$. The initial
guess for \eqref{eq:subopt2} at iteration $k$, denoted as $U_k^{(0)}$,
is the convergent orthonormal matrix $U_{k-1}$ from previous iteration
with a small random perturbation, \latin{i.e.},
\begin{equation}
    U_k^{(0)} = \orth{U_{k-1} + \mathrm{rand}(\morb, \norb)},
\end{equation}
where $\mathrm{rand}(\morb, \norb)$ denotes a random matrix of size $\morb$
by $\norb$ with each entry sampled from normal distribution with mean $0$
and standard deviation $0.1$. Using such an initial guess, the convergence is
empirically found much faster than that using a purely random initial guess.
Adding randomness to the initial guess in many cases helps with escaping
from local minima. A similar observation is obtained by the stochastic
CASSCF method~\cite{LiManni2016,LiManni2018}, where the randomness is added
to RDMs via FCIQMC. We emphasize that this is a crucial point making our
method achieve a lower ground state energy than conventional CASSCF methods.

\subsubsection*{Convergence}

We first discuss the convergence of solving \eqref{eq:subopt1} and
\eqref{eq:subopt2} and then move to the discussion on the convergence
of OptOrbFCI.

The convergence of CDFCI algorithm in solving \eqref{eq:subopt1} is
discussed in detail in \citet{Li2019c}. Since CDFCI rewrites the linear
eigenvalue problem as an unconstrained optimization problem with a nonconvex
objective function, the global convergence is guaranteed without rate and
the local convergence with a linear rate is also proved in the
compression-free setting.

The convergence analysis of the projection method with alternating BB
stepsize is proposed in \citet{Gao2018} for solving general orthonormal
constrained optimization problems, which include our subproblem
\eqref{eq:subopt2}. This method is guaranteed to converge to points with
first-order optimality condition; \latin{i.e.}, these points have a
vanishing gradient along the tangent plane of the constraint.

The convergence analysis of OptOrbFCI has not been rigorously
shown and is beyond the scope of this paper. However, the rich
literature in the convergence analysis of the alternating direction
method of multipliers~\cite{Deng2016} and coordinate-wise descent
methods~\cite{Nesterov2012, Wright2015, Shi2016, Li2019c} sheds light
on the analysis of OptOrbFCI. In general, the convergence analysis of
the overall alternating algorithm relies on the convergence analysis
of subproblems and the property of the overall objective function.
If we apply the alternating algorithm to \eqref{eq:genopt}, since
the space of $\ket{\Phi}$ remains unchanged, the energy is guaranteed
to decrease monotonically.  Hence, if we have the equivalence between
\eqref{eq:oriopt} and \eqref{eq:genopt} for all $U$, then we also have
a monotone decreasing property for solving \eqref{eq:oriopt}. Together
with the convergence properties of both subproblems, we know that
OptOrbFCI converges to points with first-order optimality condition.

\subsubsection*{Stopping Criteria}

There are plenty choices of stopping criteria for each of three
iterative algorithms. In practice, we use the following stopping
criteria joined with a fixed maximum number of iterations.

In CDFCI, we monitor the exponential moving average of the norm of
the coefficient difference, \latin{i.e.},
\begin{equation}
    S_t = (1 - \alpha) \norm{\Delta x_t} + \alpha S_{t-1},
\end{equation}
where $t$ is the iteration index, $\alpha = 0.99$ is the decay
factor, $\Delta x_t$ denotes the coefficient difference, and $S_t$
is the moving average. CDFCI stops if $S_t$ is smaller than a given
tolerance.

The stopping criterion of the projection method for the subproblem with
fixed $U$ is similar, \latin{i.e.},
\begin{equation}
    S_t = (1 - \alpha) \abs{\Delta E_t} + \alpha S_{t-1},
\end{equation}
where $\Delta E_t$ is the difference of objective functions $P_4(U_t)$
and $P_4(U_{t-1})$, and $\alpha = 0.8$ is the decay factor. If $S_t$
is smaller than a given tolerance, we stop the projection method.

In OptOrbFCI, we observe monotone decay of the FCI energy. Hence
the algorithm stops if the per-iteration decay is smaller than a
given tolerance.

\subsubsection*{Computational Complexities}

The computational complexity for an iterative algorithm depends
on both the per-iteration complexity and the number of iterations.
Our discussion also follows these two parts.

For the CDFCI algorithm, each iteration applies the Hamiltonian operator to
a single Slater determinant. The per-iteration computational cost is
dominated by the double excitation part, which selects two electrons and
excites them to two unoccupied orbitals. Hence, CDFCI costs $\bigO(\norb^2
\nelec^2)$ operations per-iteration. However, the number of iterations is
usually big, which is still believed to be of the order $\bigO({\norb
\choose \nelec})$ with a small prefactor. In practice, the iteration number
is usually around $10^6$ to $10^8$ for small systems we have tested to
achieve $10^{-1}$ mHa accuracy. The computational complexity in producing
RDMs is similar to that of the CDFCI solver part.

For the projection method, each iteration computes the gradient of the
objective function, whose computational cost is dominated by contracting a
four-way tensor $v_{pqrs}$ with $U$ matrix in three dimensions. The
per-iteration, hence, costs $\bigO(\morb^4 \norb)$ operations. The number of
iterations is much smaller than that in CDFCI. For systems we have tested,
iteration numbers are around a few hundred to a few thousands for the first
two iterations in the overall algorithm. Starting from the third iteration,
the iteration number of the projection method quickly drops to a couple
hundreds depending on the level of random perturbation on the initial value.

Putting the computational complexity for both CDFCI and the projection
method together, we have a per-iteration cost for OptOrbFCI.  When $\morb$
is not much bigger than $\norb$, the CDFCI part dominates the computation
cost and the projection method part can be ignored. However, when $\morb$ is
much bigger than $\norb$, \latin{e.g.,} when the cc-pV5Z basis set is used,
the computational cost of the projection method is not negligible, but the
CDFCI part is still more expensive. Regarding the iteration number,
OptOrbFCI usually achieves chemical accuracy in a few iterations. The
convergence to an accuracy $10^{-2}$ mHa can also be achieved within two
dozen iterations for all the cases we have tested.

\subsection{Comparison with CASSCF Algorithms}

We compare OptOrbFCI with two conventional CASSCF algorithms,
\latin{i.e.}, the Newton-Raphson~\cite{Roos1980} and super-CI
methods~\cite{Siegbahn1980, Siegbahn1981}. In the following, we first
briefly review these two methods, in particular from an optimization
point of view, and then compare them with our proposed OptOrbFCI
algorithm.

Conventional CASSCF algorithms start with a different representation for the
orbital rotation matrix. Recall that OptOrbFCI directly deals with the
partial unitary matrix with an orthonormality constraint. While in the
CASSCF framework, the orbital rotation is given by a square unitary matrix
$U$ parametrized as
\begin{equation} \label{eq:Upara}
    U = e^X,
\end{equation}
with $X$ being a skew-symmetric matrix. We denote the Slater
determinant of orbitals $\{\psi_1, \dots, \psi_\norb\}$ as
$\{\ket{D_i}\}$, so a wave function $\ket{\Psi} \in \calD[(\psi_1,
\dots, \psi_\norb)]$ \footnote{Recall that $\calD[(\psi_1, \dots,
\psi_\norb)]$ is the space spanned by Slater determinants given by
$\{\psi_1, \dots, \psi_{\norb}\}$.} can be written as
\begin{equation} \label{eq:linearcombx}
    \ket{\Psi} = \sum_{i \in \calI} x_i \ket{D_i} ,
\end{equation}
where $x_i$ are linear combination coefficients and $\calI$ denotes
the set of all configurations out of $\norb$ orbitals.  The target
wave function after rotation is then given by
\begin{equation} \label{eq:PhiPsi}
    \ket{\Phi} = \op{U} \ket{\Psi} = e^{\op{X}} \ket{\Psi},
\end{equation}
where $\op{U} = e^{\op{X}}$ denotes the rotation operator on the
Slater determinants (and hence the span) corresponding to $U =
e^X$ in \eqref{eq:Upara}.

From the point of view of optimization, the Newton-Raphson method first
converts \eqref{eq:oriopt} to an unconstrained optimization problem
using \eqref{eq:Upara} for the orbital rotation matrix, given by
\begin{equation}\label{eq:nropt}
    \min_{X, \{x_i\}: \norm{x} = 1} E \bigl(X, \{x_i\}\bigr)
\end{equation}
with
\begin{equation}
    E\bigl(X, \{x_i\}\bigr)  = \ev**{e^{-\op{X}}\oriHop
    e^{\op{X}}}{\Psi},
\end{equation}
where $\Psi$ is given by \eqref{eq:linearcombx}, so that $\norm{x} = 1$ is
equivalent to the normality constraint for $\ket{\Phi}$ due to the
orthonormality between Slater determinants.  Note that with fixing $X$, the
optimization of $E$ with respect to $\{x_i\}$ leads to a standard eigenvalue
problem, hence the exact optimum can be obtained via FCI solvers, similar to
OptOrbFCI. The optimization with respect to $X$ becomes unconstrained, so
the standard second order optimization method can be applied. However, as a
price to pay, the dependence of $E$ on $X$ becomes quite complicated due to
the parametrization~\eqref{eq:Upara}. In the Newton-Raphson method, one
approximates $E(X)$ quadratically near $X=0$. The optimization of $X$ using
the surrogate quadratic approximation leads to the linear system
\begin{equation} \label{eq:newton}
    \frac{\partial E}{\partial X_{pq}} \bigg\vert_0
    + \sum_{r < s} \frac{\partial^2 E}{\partial X_{pq} \partial
    X_{rs}} \bigg\vert_0 X_{rs} = 0\,.
\end{equation}
To write down the equation more explicitly, let us introduce a
short-hand notation for the singly-excited state as
\begin{equation}
    \ket{pq} = (\op{c}_p^\dagger \op{c}_q - \op{c}_q^\dagger \op{c}_p)
    \ket{\Psi}.
\end{equation}
Then the first order derivative at $X=0$ reads
\begin{equation}
    \frac{\partial E}{\partial X_{pq}} \bigg\vert_0 = 2
    \mel**{\Psi}{\oriHop}{pq},
\end{equation}
and the second order derivative reads
\begin{equation} \label{eq:secondderivative}
    \begin{aligned}
        \frac{\partial^2 E}{\partial X_{pq} \partial X_{rs}}
        \bigg\vert_0 & = 2 \mel**{pq}{\oriHop}{rs} \\
        & \qquad + \mel**{pq}{ (\op{c}_s^\dagger \op{c}_r -
        \op{c}_r^\dagger \op{c}_s) \oriHop}{\Psi} \\
        & \qquad + \mel**{rs}{ (\op{c}_q^\dagger \op{c}_p -
        \op{c}_p^\dagger \op{c}_q) \oriHop}{\Psi}.
    \end{aligned}
\end{equation}
After $X$ is obtained in each macro iteration, the orbitals are rotated
based on $X$~\cite{Banerjee1976, Roos1980, Siegbahn1980}. When the exact
Hessian is used and the rotation based on $X$ is handled carefully (so that
it is at least second order accurate for small $X$), the Newton-Raphson
method has local quadratic convergence~\cite{Banerjee1976}.

The super-CI method takes a slightly different point of view by directly
taking an expansion of \eqref{eq:PhiPsi} (instead of $E$) with respect to
$X$. The first order approximation of $e^{\op{X}} \ket{\Psi}$ is known as
the singly-excited wave function as
\begin{equation}
    \ket{\Psi_{\textrm{SCI}}} = \ket{\Psi} + \sum_{r<s} X_{rs}
    (\op{c}_r^\dagger \op{c}_s - \op{c}_s^\dagger \op{c}_r)\ket{\Psi},
\end{equation}
where the subscript SCI is short for singly-excited CI. To determine
$X$, the energy of $\ket{\Psi_{\textrm{SCI}}}$ is minimized; as it
is not necessarily normalized, we minimize the Ritz value
\begin{equation*}
    \frac{\ev**{\oriHop}{\Psi_{\textrm{SCI}}}}
    {\braket{\Psi_{\textrm{SCI}}}{\Psi_{\textrm{SCI}}}}
\end{equation*}
with respect to $X$, which is equivalent to solving the eigenvalue
problem of the matrix
\begin{equation}
    \begin{bmatrix}
        \ev**{\oriHop}{\Psi} & \mel**{\Psi}{\oriHop}{rs} \\
        \mel**{pq}{\oriHop}{\Psi} & \mel**{pq}{\oriHop}{rs}
    \end{bmatrix},
\end{equation}
where the second column and second row are block matrices index by $rs$
and $pq$, respectively. Thus, each step of super-CI can also be viewed
as solving an eigenvalue problem in an extended variational space.
Compared with the Newton-Raphson method, the matrix above is related
to the Hessian used in the Newton-Raphson method \eqref{eq:newton}. The
last two terms in the second derivative \eqref{eq:secondderivative}
are missing in the super-CI matrix, due to the different approximation
taken in the expansion.

\smallskip 

In both CASSCF algorithms, the rotation of the orbitals according to $X$
needs to be processed very carefully.  Direct transformation using the first
order approximation of \eqref{eq:Upara} is manageable if orbitals are then
orthogonalized or an overlapping matrix is introduced. An alternative
approach is through the natural orbital of the singly excited wave function
$\ket{\Psi_{SCI}}$.

We emphasize that not every element of $X$ is involved in the above
calculation. Since the energy $E$ is invariant to the rotation within
unselected orbitals, the elements $X_{pq}$ for both $p$ and $q$
corresponding to unselected orbitals are ignored, which also improves the
numerical stability of the above algorithms. Moreover the energy $E$ is also
invariant to the rotation within selected orbitals.  If the direct FCI
solver is applied, the elements $X_{pq}$ for both $p$ and $q$ corresponding
to selected orbitals can be ignored as well, while, if modern FCI solvers
are applied, which all include some compression of the coefficients, the
rotation within the selected orbitals often helps improve the
compressibility of wave function coefficient; hence, they are preserved in
the calculations~\cite{Tubman2018a, Levine2020}. In the end, the numbers of
degrees of freedom in all three algorithms are the same.

Recall that the energy is only a fourth order polynomial of the unitary
matrix $U$ as shown in \eqref{eq:subobj2}, while on the other hand, after
introducing the parametrization \eqref{eq:Upara}, the energy depends in a
quite complicated way on the parameter matrix $X$. Conventional CASSCF
algorithms then introduce approximations to $E$ and $\op{U} \ket{\Psi}$.
Expressions are valid when $X$ is around zero, which means that $U$ is close
to an identity matrix. Hence, at each macro step, conventional CASSCF
algorithms are valid and efficient if the rotation of orbitals is not far
from identity. There are two potential drawbacks of this local optimization:
1) many macro iterations are needed to move the rotation matrix away from
its initialization; 2) algorithms converge efficiently to a local minimum
close to the initial value.  In comparison, OptOrbFCI adopts modern
optimization techniques for orthonormal constrained optimization problems
and is free to converge to any orthonormal matrix in each macro iteration.
Therefore, each orbital optimization problem is solved more accurately and
the algorithm potentially converges to better minima with lower energies.
Specifically, taking a random initial unitary matrix is feasible in
OptOrbFCI, while it leads to unsatisfactory results in conventional CASSCF
calculations.  The price to pay is possibly a more expensive orbital
optimization cost compared with conventional CASSCF algorithms. However, we
find that such a cost is negligible compared to the cost of FCI solvers,
which is the setting that motivates our work. 

\begin{remark}
    In CASSCF, the orbitals are usually split into three groups, inactive,
    active, and virtual. Active and virtual orbitals correspond to the
    selected orbitals and unselected ones after rotation. Inactive orbitals
    are orbitals frozen to be doubly occupied ones. Introducing the inactive
    orbitals does not change the structure of any optimization algorithm
    above. With another set of indices denoting the inactive orbitals, many
    matrix/tensor elements are zeros, which help reduce the computational
    cost. We omit the related expressions for simplicity.
\end{remark}

\section{Numerical Results}
\label{sec:numerical results}

In this section, we demonstrate the efficiency of the proposed OptOrbFCI
through several numerical experiments. First, we explore the detailed
properties of OptOrbFCI through a sequence of numerical experiments on a
single water molecule. A comparison against the CASSCF method is explored
here as well. Then we compare the ground-state energies of the carbon dimer
and nitrogen dimer calculated through OptOrbFCI under various basis sets,
\latin{i.e.}, cc-pVDZ, cc-pVTZ, cc-pVQZ, and cc-pV5Z. Finally, we adopt
OptOrbFCI to benchmark the binding curve of the nitrogen dimer under the
cc-pVQZ basis set, which consists of systems with various levels of
correlations. And the dissociation energy for the nitrogen dimer is also
compared against that through the FCI method under various basis sets.

In all the numerical experiments, the original given orbitals (one-body and
two-body integrals) are calculated via the restricted HF (RHF) in
PSI4~\cite{PSI4} package. All energies are reported in the unit Hartree
(Ha).

We adopt the modern C++ implementation of CDFCI~\cite{CDFCI} and our own
version of the projection method~\cite{Gao2018} implemented in MATLAB.
Multithread parallelization is disabled in CDFCI. The communication between
CDFCI and the projection method is done via file system, \latin{i.e.}, the
FCIDUMP file and RDM files.  All results labeled by FCI are produced by
CDFCI.  The implementation of the CASSCF method in PySCF~1.7.1~\cite{pyscf}
is applied for comparison purposes.

\subsection{\ch{H2O} Molecule}

The water molecule used in this section is at its equilibrium
geometry~\cite{Booth2009, Wang2019}, \latin{i.e.}, OH bond
length $1.84345$ $a_0$ and HOH bond angle $110.6$\textdegree.
Table~\ref{tab:h2obasisset} summarizes the properties associated with
different basis sets.

\begin{table*}[h]
    \centering
    \begin{tabular}{cccccc}
        \toprule
        Molecule & Basis   & Electrons & Orbitals &
        HF energy & GS energy \\
        \toprule
        \multirow{4}{*}{\ch{H2O}}
        & cc-pVDZ & 10 &  24 & $-76.0240386$ & $-76.2418601$ \\
        & cc-pVTZ & 10 &  58 & $-76.0544374$ & -- \\
        & cc-pVQZ & 10 & 115 & $-76.0621073$ & -- \\
        & cc-pV5Z & 10 & 201 & $-76.0644002$ & -- \\
        \bottomrule
    \end{tabular}
    \caption{Basis Sets for \ch{H2O}. HF energy denotes the
    Hartree-Fock energy calculated by PSI4~\cite{PSI4} and GS
    energy denotes the FCI ground-state energy calculated by
    CDFCI~\cite{Wang2019}. A bar means the number is not available.}
    \label{tab:h2obasisset}
\end{table*}

For CDFCI, the compression threshold is $5 \times 10^{-7}$, the
tolerance for convergence is $5 \times 10^{-6}$, and the maximum number
of iterations is $3 \times 10^{7}$. The convergence tolerance for the
projection method is $10^{-7}$, and the maximum number of iterations
is $10^4$. For OptOrbFCI, the convergence tolerance is $10^{-4}$
and the maximum number of iterations is $20$. These settings are used
for all numerical experiments of \ch{H2O} molecule.

\begin{figure}[t]
    \centering
    \includegraphics[width=0.5\textwidth]{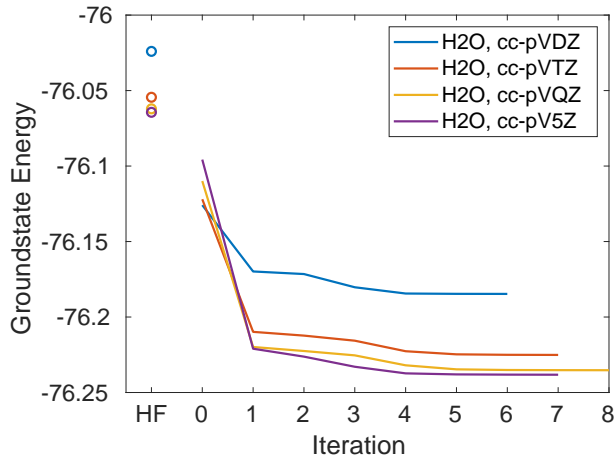}
    \caption{Convergence of the ground-state energy of \ch{H2O} against
    iteration for $\norb = 12$.} \label{fig:h2o-12}
\end{figure}

\begin{figure}[t]
    \centering
    \includegraphics[width=0.5\textwidth]{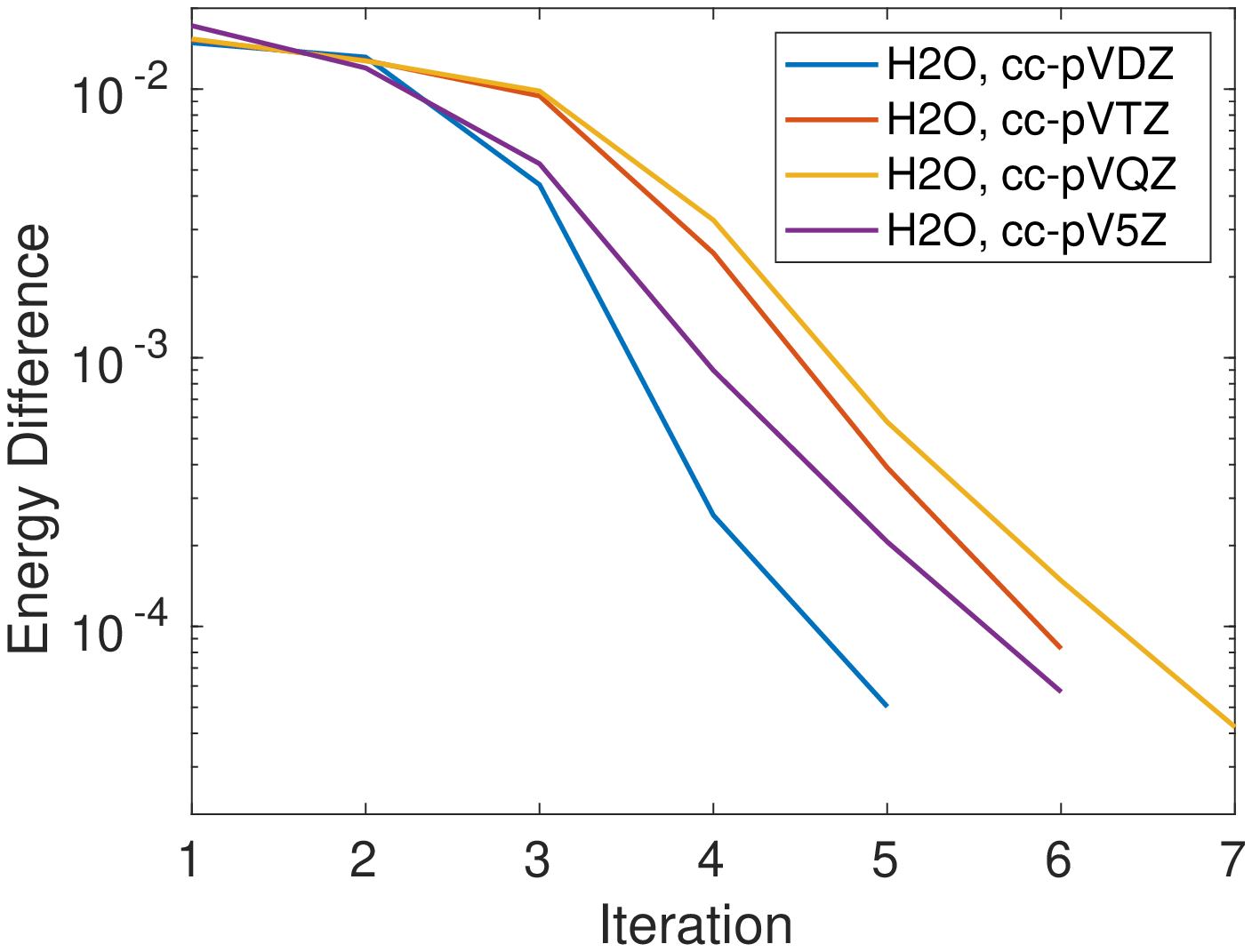}
    \caption{Difference of the ground-state energy of \ch{H2O} against
    iteration for $\norb = 12$.} \label{fig:h2o-12-diff}
\end{figure}

\begin{figure}[t]
    \centering
    \includegraphics[width=0.5\textwidth]{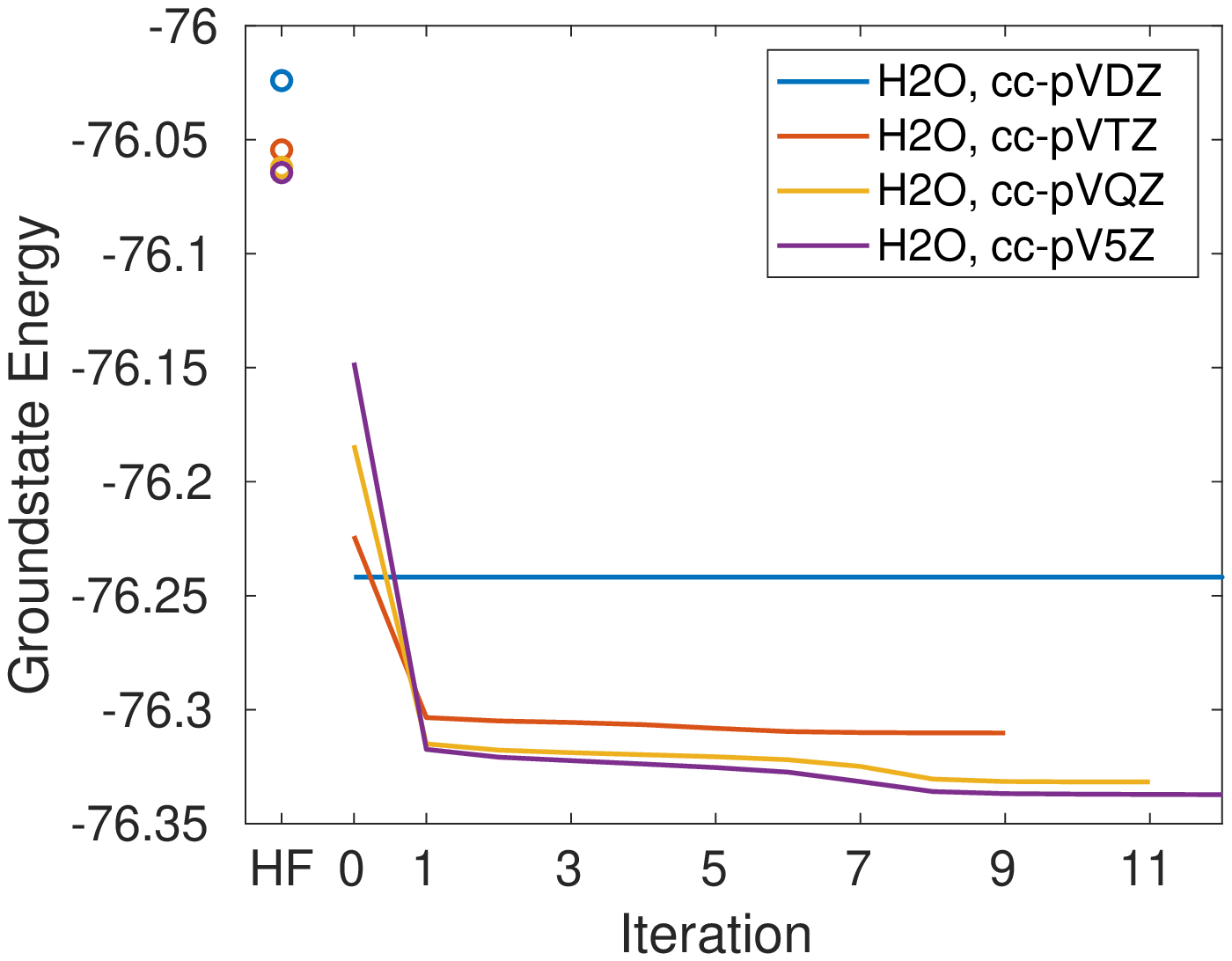}
    \caption{Convergence of the ground-state energy of \ch{H2O} against
    iteration for $\norb = 24$.} \label{fig:h2o-24}
\end{figure}

\begin{figure}[t]
    \centering
    \includegraphics[width=0.5\textwidth]{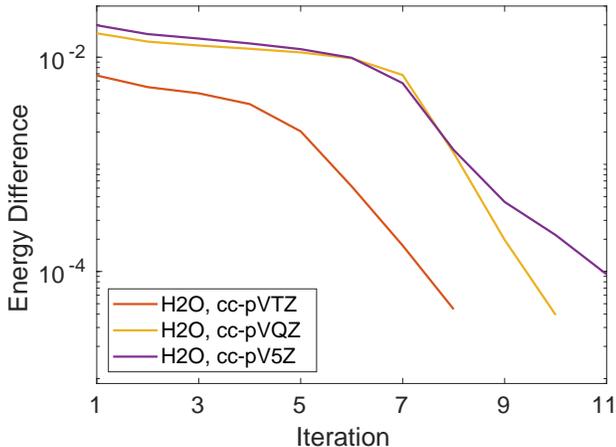}
    \caption{Difference of the ground-state energy of \ch{H2O} against
    iteration for $\norb = 24$.} \label{fig:h2o-24-diff}
\end{figure}

Two different numbers of selected orbitals, $\norb = 12$ and
$\norb = 24$, are tested for \ch{H2O} molecules on a sequence of
basis sets.  Figure~\ref{fig:h2o-12} and Figure~\ref{fig:h2o-24}
show the convergence behavior of OptOrbFCI against the iteration
number for $\norb = 12$ and $\norb = 24$ respectively. The HF
energies are also plotted in both figures with the $x$-axis label
being ``HF''. The energies associated with iteration $0$ is the FCI
energies before applying projection method and the orbitals with
smallest $\norb$ orbital energies are used as the selected orbitals.
Figure~\ref{fig:h2o-12-diff} and Figure~\ref{fig:h2o-24-diff} further
show the log scale of the energy difference against the iteration. Here
the energy difference is defined as the difference between the
FCI ground-state energy at current iteration and the converged FCI
ground-state energy.  In Figure~\ref{fig:h2o-24-diff}, the curve
associated with cc-pVDZ is removed since the ground-state energies
stay constant throughout iterations. Table~\ref{tab:h2o-energy}
lists all convergent FCI ground-state energies.

\begin{table}
    \centering
    \begin{tabular}{ccc}
        \toprule
        & $\norb = 12$ & $\norb = 24$ \\
        Basis   & GS energy & GS energy \\
        \toprule
        cc-pVDZ & $-76.1846948$ & $-76.2418601$ \\
        cc-pVTZ & $-76.2251082$ & $-76.3102225$ \\
        cc-pVQZ & $-76.2352354$ & $-76.3317350$ \\
        cc-pV5Z & $-76.2382165$ & $-76.3372849$ \\
        \bottomrule
    \end{tabular}
    \caption{Ground-state energies for \ch{H2O} with different
    number of selected orbitals under variant basis sets.}
    \label{tab:h2o-energy}
\end{table}

In both Figure~\ref{fig:h2o-12} and Figure~\ref{fig:h2o-24}, we notice
that all FCI ground-state energies are lower than HF energy under any
these basis set.  For the first FCI calculation with selected orbitals
according to lowest orbital energies, \latin{i.e.}, iteration $0$, we
observe that the smaller the basis set the lower the energy. This is
likely due to the energy concentration of orbitals, which means that
smaller basis set has better concentration of energies among occupied
orbitals. As long as an optimized partial unitary matrix $U$ is applied,
such an order no longer preserves starting from iteration $1$. In both
cases, we also notice that the ordering of energies for different basis
sets reveals after the first two iterations. Starting from then, larger
basis sets consistently have lower ground-state energies than the
smaller basis sets. The difference between the ground-state energies for
different basis sets are much larger than the desired chemical accuracy.
Further in Figure~\ref{fig:h2o-12-diff} and
Figure~\ref{fig:h2o-24-diff}, steady convergence is observed for all
experiments and OptOrbFCI converges to chemical accuracy level within a
few iterations. Larger $\norb$ leads to slightly more iterations in
OptOrbFCI.

In addition to Figure~\ref{fig:h2o-12} and Figure~\ref{fig:h2o-24},
Table~\ref{tab:h2o-energy} further illustrates ground-state energies
for both $\norb = 12$ and $\norb = 24$. The difference between
neighbour basis sets is decreasing as the basis set size increases.
The decrease of energies from cc-pVQZ to cc-pV5Z for both $\norb$
are on the level of millihartree. Hence the basis limit is nearly
achieved for \ch{H2O} given $\norb = 12$ and $\norb = 24$.

\begin{figure}[ht]
    \centering
    \includegraphics[width=0.5\textwidth]{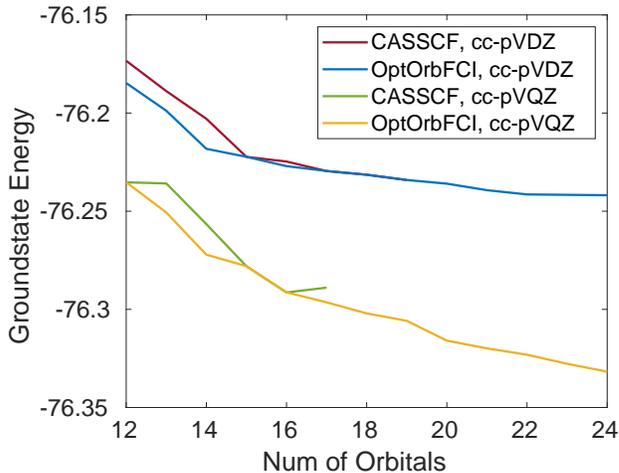}
    \caption{Convergence of the ground-state energy of \ch{H2O} against
    varying $\norb$.} \label{fig:h2o-orbital}
\end{figure}

\begin{table}[h]
    \centering
    \begin{tabular}{ccccc}
        \toprule
        & \multicolumn{2}{c}{OptOrbFCI} & \multicolumn{2}{c}{CASSCF} \\
        \cmidrule(lr){2-3}
        \cmidrule(lr){4-5}
        Orbs & GS energy & Iter & GS energy & Iter \\
        \toprule
        \rowcolor{lightgray}
        12 & $-76.1847$ &  6 & $-76.1734$ & 7 \\
        \rowcolor{lightgray}
        13 & $-76.1988$ &  8 & $-76.1888$ & 7 \\
        \rowcolor{lightgray}
        14 & $-76.2182$ &  8 & $-76.2029$ & 7 \\
        15 & $-76.2223$ &  7 & $-76.2223$ & 7 \\
        \rowcolor{lightgray}
        16 & $-76.2270$ & 10 & $-76.2247$ & 7 \\
        17 & $-76.2295$ &  6 & $-76.2295$ & 7 \\
        18 & $-76.2314$ &  6 & $-76.2314$ & 6 \\
        19 & $-76.2341$ &  3 & $-76.2341$ & 5 \\
        20 & $-76.2360$ &  3 & $-76.2360$ & 4 \\
        \bottomrule
    \end{tabular}
    \caption{Comparison of OptOrbFCI and CASSCF~\cite{pyscf} for
    \ch{H2O} under cc-pVDZ basis set. Gray row indicates significantly
    different ground-state energies for different methods. Orbs is
    the number of selected orbitals and Iter is the macro iteration
    number.} \label{tab:h2o-comp-ccpvdz}
\end{table}

\begin{table}[h]
    \centering
    \begin{tabular}{ccccc}
        \toprule
        & \multicolumn{2}{c}{OptOrbFCI} & \multicolumn{2}{c}{CASSCF} \\
        \cmidrule(lr){2-3}
        \cmidrule(lr){4-5}
        Orbs & GS energy & Iter & GS energy & Iter \\
        \toprule
        12 & $-76.2352$ &  8 & $-76.2353$ & 19 \\
        \rowcolor{lightgray}
        13 & $-76.2506$ &  7 & $-76.2358$ &  6 \\
        \rowcolor{lightgray}
        14 & $-76.2721$ &  9 & $-76.2566$ &  6 \\
        15 & $-76.2780$ &  5 & $-76.2780$ &  6 \\
        16 & $-76.2913$ & 15 & $-76.2914$ & 19 \\
        \rowcolor{lightgray}
        17 & $-76.2964$ & 18 & $-76.2889$ &  8 \\
        \bottomrule
    \end{tabular}
    \caption{Comparison of OptOrbFCI and CASSCF~\cite{pyscf} for
    \ch{H2O} under cc-pVQZ basis set.}
    \label{tab:h2o-comp-ccpvqz}
\end{table}

The decrease of the energy as $\norb$ increases from $12$ to $24$ is
still significant for all basis sets. Hence we further investigate the
relationship between the ground-state energy and the number of
selected orbitals, $\norb$. Figure~\ref{fig:h2o-orbital} shows such a
relationship under cc-pVDZ and cc-pVQZ basis sets. As shown in
Figure~\ref{fig:h2o-orbital}, as we gradually increase the number of
selected orbitals, the ground-state energy of cc-pVDZ basis set first
decay rapidly for $\norb$ between $12$ to $15$, and then, for
$\norb \geq 15$, the decay is much slower. The decay of the
ground-state energy of cc-pVQZ basis set decreasing steadily for all
$\norb$ tested here. Hence we expect the slow decay for cc-pVQZ basis
set comes later than $\norb = 24$. While, under limited computational
budget, the ground-state energy for cc-pVQZ with $24$ selected
orbitals is already much lower than that of cc-pVDZ with $24$ selected
orbitals.

In addition to Figure~\ref{fig:h2o-orbital}, the comparison between
OptOrbFCI and CASSCF is detailed in Table~\ref{tab:h2o-comp-ccpvdz}
and Table~\ref{tab:h2o-comp-ccpvqz} for cc-pVDZ and cc-pVQZ basis sets
respectively.  In both tables, we highlight the rows with significantly
different ground-state energies. In all cases, OptOrbFCI achieves
lower energy. Since the original optimization problem~\eqref{eq:oriopt}
is non-convex, any method could be trapped in local minima especially
for methods concerning local optimization. OptOrbFCI, using additive
random perturbation to initializations in orbital optimization,
in many cases avoids the local minima near the initial point. Hence
we observe that OptOrbFCI in many cases achieves lower ground-state
energy and in no case achieves higher ground-state energy. Here both
methods use the same default initial one- and two-body integrals
with respect to Hatree-Fock orbitals. When different initializations
are considered, the results in Table~\ref{tab:h2o-comp-ccpvdz} and
Table~\ref{tab:h2o-comp-ccpvqz} could be different. While OptOrbFCI
is still expected to achieve energies lower or equal to that of
CASSCF. If we further compare the macro iteration numbers, when
both methods converge to the same ground-state energy, OptOrbFCI
has less or equal number of macro iterations comparing to CASSCF.
Even for those cases where lower ground-state energy is achieved by
OptOrbFCI, the difference in macro iteration number is, in most cases,
not significant.  Hence we conclude that OptOrbFCI could achieve
lower ground-state energy and reduce the macro iteration number.

\subsection{\ch{C2} and \ch{N2}}

This section studies OptOrbFCI applied to \ch{C2} and \ch{N2} under their
equilibrium geometry; \latin{i.e.,} the bond length for \ch{C2} is $1.24253$
\r{A}~\cite{Holmes2016, Wang2019} and the bond length for \ch{N2} is $2.118$
$a_0$~\cite{Chan2004, Wang2019}.

The hyper parameters in OptOrbFCI are the same for \ch{C2} and
\ch{N2}.  In CDFCI, the compression threshold is $5 \times 10^{-6}$,
the tolerance for convergence is $10^{-5}$, and the maximum number
of iterations is $3 \times 10^{7}$.  In the projection method,
the convergence tolerance is $10^{-7}$ and the maximum number of
iterations is $10^{4}$. In OptOrbFCI, the convergence tolerance is
$10^{-4}$ and the maximum number of iterations is $20$.

\begin{table*}[h]
    \centering
    \begin{tabular}{cccccccc}
        \toprule
        &&&& & Selected & Iteration & OptOrbFCI \\
        Molecule & Basis & Electrons & Orbitals & HF energy
        & Orbitals & Number & GS energy \\
        \toprule
        \multirow{4}{*}{\ch{C2}}
        & cc-pVDZ & 12 &  28 & $-75.4168820$ & 28 &  - & $-75.7319604$ \\
        & cc-pVTZ & 12 &  60 & $-75.4014464$ & 28 &  6 & $-75.7763001$ \\
        & cc-pVQZ & 12 & 110 & $-75.4057650$ & 28 & 10 & $-75.7991578$ \\
        & cc-pV5Z & 12 & 182 & $-75.4065236$ & 28 & 12 & $-75.8030425$ \\
        \bottomrule
    \end{tabular}
    \caption{Basis sets and numerical results for \ch{C2}.}
    \label{tab:c2}
\end{table*}

\begin{table*}[h]
    \centering
    \begin{tabular}{cccccccc}
        \toprule
        &&&& & Selected & Iteration & OptOrbFCI \\
        Molecule & Basis & Electrons & Orbitals & HF energy
        & Orbitals & Number & GS energy \\
        \toprule
        \multirow{4}{*}{\ch{N2}}
        & cc-pVDZ & 14 &  28 & $-108.9493779$ & 28 &  - & $-109.2821727$ \\
        & cc-pVTZ & 14 &  60 & $-108.9775136$ & 28 &  7 & $-109.3409252$ \\
        & cc-pVQZ & 14 & 110 & $-108.9849510$ & 28 &  7 & $-109.3639435$ \\
        & cc-pV5Z & 14 & 182 & $-108.9866093$ & 28 & 13 & $-109.3689430$ \\
        \bottomrule
    \end{tabular}
    \caption{Basis sets and numerical results for \ch{N2}.}
    \label{tab:n2}
\end{table*}

Table~\ref{tab:c2} and Table~\ref{tab:n2}, for \ch{C2} and \ch{N2}
respectively, show the properties of the dimers and our numerical
results. Since OptOrbFCI selects the number of orbitals the same
as that under the cc-pVDZ basis set, the ground-state energies of
cc-pVDZ basis set are the FCI results and are used as reference for the
rest results. Similar figures as in the case of \ch{H2O} can also be
plotted for \ch{C2} and \ch{N2}. Since there is not much difference,
we omit them from the paper.

Both Table~\ref{tab:c2} and Table~\ref{tab:n2} show similar properties and
we discuss their numerical results together. First of all, we notice that
any FCI ground-state energy is lower than all the HF energies, which shows
that the improvement of the FCI calculation over the HF calculation is
beyond the difference between basis sets. Since we fix the number of
selected orbitals to the same as that under the cc-pVDZ basis set, the
computational cost of the optimal orbital selection method for other basis
sets remains the same order as the cost of FCI under the cc-pVDZ basis set.
If only the ground-state energy is needed, then OptOrbFCI is roughly twice
the iteration number more expensive then that of the FCI under cc-pVDZ. If
both the ground-state energy and the RDMs are needed for downstream tasks,
then the increasing factor is reduced to the iteration number, which is
between $6$ and $13$. In these estimations, the computational cost of the
projection method is ignored. This is the case for the cc-pVTZ and cc-pVQZ
basis sets, while for the cc-pV5Z basis set, the computational cost of the
projection method is still smaller than that of CDFCI part but of the same
order. Now we provide a few numbers to support this. All the numerical
results in this section are performed on a machine with Intel Xeon CPU
E5-2687W v3 at 3.10 GHz and 500 GB memory. At least $6$ tasks are performed
simultaneously. The memory for each problem is limited to 40 GB.  Given
$\norb$ selected orbitals, for all basis sets, each CDFCI part (FCI solver
plus RDM calculations) costs varying from $10,000$ to $50,000$ seconds for
\ch{C2}, while the computational costs for the projection method parts are
dramatically different for different basis sets. The projection method part
costs nearly $200$ seconds, $3,000$ seconds, and $10,000$ seconds for the
cc-pVTZ, cc-pVQZ, and cc-pV5Z basis sets respectively. The runtime for
\ch{N2} has a similar ratio between the CDFCI part and the projection
method.

Comparing the ground-state energies under different basis sets, we
notice that the lower ground-state energy is achieved under the larger
basis set. The improvement between consecutive basis sets, however,
is gradually decreasing, close to exponential decay.  For both \ch{C2}
and \ch{N2}, the improvement between the cc-pV5Z and cc-pVQZ basis sets
is on the level of millihartree.

\subsection{\ch{N2} Binding Curve}

This section benchmarks the binding curve of \ch{N2} under the cc-pVQZ basis
set with $\norb = 28$, which is the number of orbitals under the cc-pVDZ
basis set. The all-electron \ch{N2} binding curve is well-known to be a
difficult problem due to the multireference property for geometry away from
equilibrium. In \citet{Wang2019}, the binding curve on a very fine grid is
produced under the cc-pVDZ basis set up to $10^{-3}$ mHa accuracy. Here we
rebenchmark the binding curve under the cc-pVQZ basis set with $\norb = 28$
selected orbitals with an accuracy up to $10^{-1}$ mHa. Since the number of
orbitals remains the same, the computational cost of our optimal orbital
selection is of the same order as a single CDFCI execution~\cite{Wang2019}.

For the binding curve, exact same geometries as in \citet{Wang2019} are
produced. The compression threshold, for the CDFCI part, is $5 \times
10^{-6}$, the tolerance for convergence is $10^{-5}$, and the maximum number
of iterations is $3 \times 10^{7}$. The convergence tolerance for the
projection method is $10^{-7}$ and the maximum number of iterations is
$10^4$. For OptOrbFCI, the convergence tolerance is $10^{-4}$ and the
maximum number of iterations is $20$.

\begin{figure}[ht]
    \centering
    \includegraphics[width=0.5\textwidth]{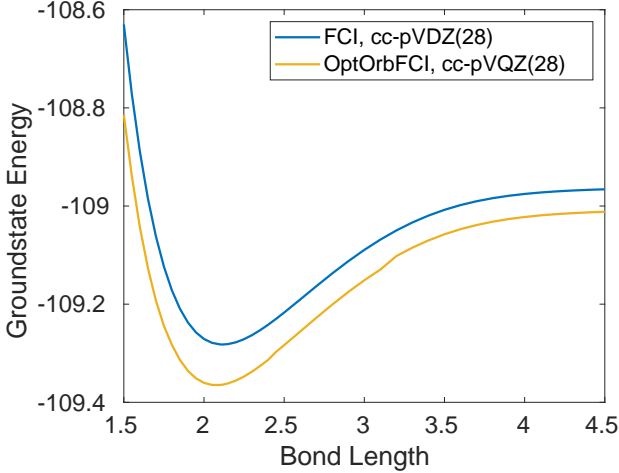}
    \caption{Binding curves for \ch{N2}. The blue curve is cited from
    CDFCI~\cite{Wang2019}. For each bond length, OptOrbFCI selects $28$
    orbitals under cc-pVQZ basis set.}\label{fig:n2-bindingcurve}
\end{figure}

\begin{figure}[ht]
    \centering
    \includegraphics[width=0.5\textwidth]{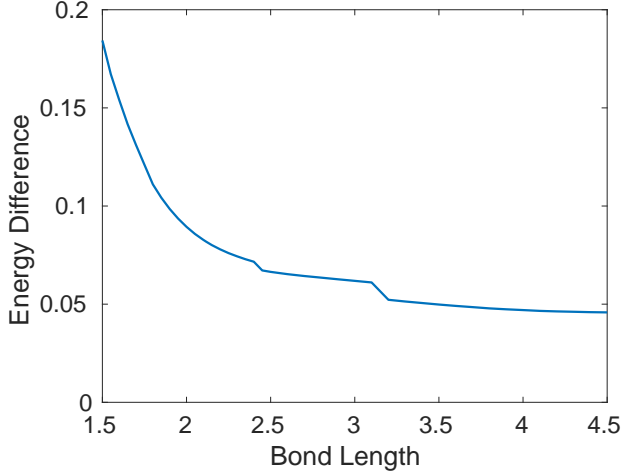}
    \caption{Difference of binding curves for \ch{N2} using FCI
    under cc-PVDZ basis set and OptOrbFCI under cc-pVQZ basis set
    with $\norb = 28$.} \label{fig:n2-bindingcurvediff}
\end{figure}

\begin{table*}[h]
    \centering
    \begin{tabular}{ccccccc}
        \toprule
        &&&& GS energy & GS energy & Dissociation \\
        Method & Basis & Electrons & Orbitals & 2.118 $a_0$
        & 4.5 $a_0$ & energy \\
        \toprule
        \multirow{2}{*}{FCI}
        & cc-pVDZ & 14 &  28 &
        $-109.2821727$ & $-108.9659102$ & $0.3162625$ \\
        & cc-pVQZ & 14 & 110 &
        $-109.4590412$ & $-109.1059938$ & $0.3530474$ \\
        OptOrbFCI
        & cc-pVQZ & 14 &  28 &
        $-109.3639435$ & $-109.0117220$ & $0.3522214$ \\
        \bottomrule
    \end{tabular}
    \caption{Dissociation energy for \ch{N2}. FCI results are
    calculated through CDFCI.} \label{tab:n2-dissociation}
\end{table*}

Figure~\ref{fig:n2-bindingcurve} illustrates the binding curves of
\ch{N2} calculated from CDFCI under cc-pVDZ basis set~\cite{Wang2019}
and from OptOrbFCI under cc-pVQZ basis set with $\norb = 28$. From
the figure, in all geometries, OptOrbFCI provides lower variational
ground-state energies, while the overall shapes for two curves remain
similar.  Figure~\ref{fig:n2-bindingcurvediff} further shows the energy
difference of two binding curves, \latin{i.e.}, the ground-state energy
of CDFCI minus that of OptOrbFCI.  We observe that the decrease is more
dramatic when two atoms are closer. There are two non-smooth points in
the energy difference around $2.45 a_0$ and $3.2 a_0$. Numerically,
we also find that the computation is more difficult around these two
bond lengths, \latin{i.e.}, the number of iterations increases. Further
investigation is needed around these two points.

Comparing to the single ground-state energy, the energy gap is of
more chemical relevance. Here, we also include the dissociation
energies for $\ch{N2}$ under three settings. The dissociation
energy is defined as the difference of ground-state energies at
equilibrium geometry (2.118 $a_0$) and at well separated geometry
(4.5 $a_0$). Three settings are FCI under cc-pVDZ, FCI under cc-pVQZ,
and OptOrbFCI under cc-pVQZ with $\norb = 28$. Numerical results are
listed in Table~\ref{tab:n2-dissociation}.  Using the dissociation
energy of FCI under cc-pVQZ as a reference solution, we notice that
the dissociation energy of OptOrbFCI is more accurate than that of FCI
under cc-pVDZ.  The error for FCI under cc-pVDZ is about $4 \times
10^{-2}$ Ha whereas the error for OptOrbFCI is about $10^{-3}$,
which is on the level of chemical accuracy.  Hence we conclude
that OptOrbFCI, in addition to provide lower ground-state energies,
provides more accurate dissociation energy.

\section{Conclusion and Discussion}
\label{sec:conclusion}

We consider the question in this paper for full configuration interaction
(FCI) pursuing the basis set limit under a computational budget. We propose
a coupled optimization problem~\eqref{eq:oriopt} as a solution to the
question, which is also the formula for CASSCF. The coupling therein between
the ground-state wave function $\ket{\Phi}$ and the partial unitary matrix
$U$ is complicated. Due to the complication, the optimization
problem~\eqref{eq:oriopt} is then split into two subproblems,
\eqref{eq:subopt1} and \eqref{eq:subopt2}, where the former is a standard
FCI problem under compressed orbitals and the latter is an optimization of a
4~th order polynomial of $U$ with orthonormality constraint. An overall
alternating iterative algorithm is proposed to address the optimization
problem~\eqref{eq:oriopt} with the first subproblem~\eqref{eq:subopt1}
solved by a wave function based FCI solver, namely CDFCI~\cite{Wang2019} and
the second subproblem~\eqref{eq:subopt2} solved by a projection
method~\cite{Gao2018}. The overall method above is referred as OptOrbFCI.
The method in general is efficient and stable. OptOrbFCI usually converges
in 5 to 15 iterations to achieve up to $10^{-1}$ mHa accuracy. The
computational cost, hence, is bounded by that of a few executions of the FCI
solver on the selected orbital sets.

Numerically, we apply OptOrbFCI to the water molecule, carbon dimer, and
nitrogen dimer under variant basis sets.  Under the number of orbitals using
the cc-pVDZ basis set, we pursue the FCI calculation under cc-pVTZ, cc-pVQZ,
and cc-pV5Z basis sets.  In all cases, we obtain ground-state energies lower
than that under cc-pVDZ, where the decrease is beyond chemical accuracy. In
the comparison against the conventional CASSCF method~\cite{pyscf},
OptOrbFCI could achieve lower ground-state energy and reduce the macro
iteration number. \ch{N2} binding curve is rebenchmarked using OptOrbFCI
under the cc-pVQZ basis set with $28$ selected orbitals. And the
dissociation energy in this case is more accurate than that obtained by the
FCI solver under the cc-pVDZ basis set. Hence we conclude that OptOrbFCI
coupling with existing FCI solvers is able to pursue the basis set limit
under a computational budget.

There are a list of immediate future works of OptOrbFCI. In the current
implementation, the orbital symmetry in the given large orbital set is
totally ignored; so is the frozen core setting. Under the given large
orbital set with orbital symmetry, both the one-body and two-body
integrals are of sparse structure. As we ignored the symmetry and
frozen core setting, the one-body and two-body integrals of the rotated
orbitals are then dense tensors. The downstream FCI problem becomes
more expensive. Hence one future work is to implement the rotation
under an orbital symmetry constraint and frozen core setting to reduce
the cost of FCI solvers. When orbital symmetries are preserved, the
corresponding ground-state energy will be lower bounded by that of our
current algorithm. Further investigation is needed on the trade-off
between the accuracy and the computational cost. A parallelization of
the projection method becomes important when the basis set gets
large, since the computational bottleneck for the projection method
lies in the 4-way tensor contraction, which can be realized as a dense
matrix-matrix multiplication. Efficient both distributed-memory and
shared-memory parallelizations are manageable. Highly efficient
GPU acceleration can also be expected. Besides implementation,
further investigation of the convergence property is desired. And
extension to low-lying excited states calculation is also a promising
future work to be explored. When both ground-state and
low-lying excited states are considered under the OptOrbFCI framework with
a modified objective function, we expect that the optimal rotation
matrix would balance the error among states under consideration and
hence potentially provide more accurate approximation to excitation
energies than our current algorithm.

 
\begin{acknowledgement}
    The authors thank Zhe Wang for helpful discussions. The authors
    also thank Jonathon Misiewicz and Qiming Sun for constructive
    suggestions on the comparison with CASSCF. The work is supported
    in part by the US National Science Foundation under awards
    DMS-1454939 and DMS-2012286, and by the US Department of Energy
    via grant DE-SC0019449.
\end{acknowledgement}






\nocite{*}
\bibliography{reference}

\newpage
\appendix

\section{Equivalence between \eqref{eq:oriopt-oriH} and
\eqref{eq:oriopt}}
\label{app:equivalentHam}

This section provides detailed derivations for the equivalence
between \eqref{eq:oriopt-oriH} and \eqref{eq:oriopt}. The key step
is to show that \eqref{eq:cctildePhi} holds for any wave function
$\ket{\Phi}$ in $\DpsiU = \Dphi$.  Since the operators are linear
operators and the space is a linear space, it is sufficient to show that
\eqref{eq:cctildePhi} holds for all bases in $\Dphi$, \latin{i.e.},
all Slater determinants. Any Slater determinant $\ket{D_i}$ in $\Dphi$
can be written as,
\begin{equation}
    \ket{D_i} = \op{d}_{i_1}^\dagger \cdots \op{d}_{i_{\nelec}}^\dagger
    \ket{0}
\end{equation}
where $i_1, \dots, i_{\nelec}$ are the index of $\nelec$ occupied
orbitals and $\ket{0}$ denotes vacuum state.  Now we evaluate the
difference of acting $\op{c}_q$ and $\optilde{c}_q$ on such a Slater
determinant. Using the anticommutation relation \eqref{eq:anticommcd}
and \eqref{eq:anticommctilded}, the difference can be simplified as,
\begin{equation}
    \begin{split}
        & \big( \op{c}_q - \optilde{c}_q \big) \ket{D_i} \\
        = & \big( \op{c}_q - \optilde{c}_q \big) \op{d}_{i_1}^\dagger
        \cdots \op{d}_{i_{\nelec}}^\dagger \ket{0} \\
        = & \Big( \acomm{\op{c}_q}{\op{d}_{i_1}^\dagger} -
        \acomm{\optilde{c}_q}{\op{d}_{i_1}^\dagger} \Big)
        \op{d}_{i_2}^\dagger \cdots \op{d}_{i_{\nelec}}^\dagger
        \ket{0} \\
        & - \op{d}_{i_1}^\dagger \big( \op{c}_q - \optilde{c}_q \big)
        \op{d}_{i_2}^\dagger \cdots \op{d}_{i_{\nelec}}^\dagger
        \ket{0} \\
        = & - \op{d}_{i_1}^\dagger \big( \op{c}_q - \optilde{c}_q \big)
        \op{d}_{i_2}^\dagger \cdots \op{d}_{i_{\nelec}}^\dagger
        \ket{0} \\
        = & (-1)^k \op{d}_{i_1}^\dagger \cdots \op{d}_{i_k}^\dagger
        \big( \op{c}_q - \optilde{c}_q \big) \op{d}_{i_{k+1}}^\dagger
        \cdots \op{d}_{i_{\nelec}}^\dagger \ket{0} \\
        = & (-1)^{\nelec} \op{d}_{i_1}^\dagger \cdots
        \op{d}_{i_{\nelec}}^\dagger \big( \op{c}_q - \optilde{c}_q
        \big) \ket{0} \\
        = & 0,
    \end{split}
\end{equation}
where the last equality holds since the annihilation operators acting
on the vacuum state vanish.  Since any wave function $\ket{\Phi} \in \Dphi$
can be expressed as a linear combination of Slater determinants,
\latin{i.e.}, $\ket{\Phi} = \sum_{i} x_i \ket{D_i}$, where $x_i$ are
coefficients, acting the difference of $\op{c}_q$ and $\optilde{c}_q$
on it leads to,
\begin{equation}
    \big( \op{c}_q - \optilde{c}_q \big) \ket{\Phi} 
    = \sum_i x_i \big( \op{c}_q - \optilde{c}_q \big) \ket{D_i} = 0.
\end{equation}
Hence we showed that \eqref{eq:cctildePhi} holds for all $\ket{\Phi}
\in \Dphi$. The conjugate of \eqref{eq:cctildePhi} gives,
\begin{equation}
    \bra{\Phi} \op{c}_p^\dagger = \bra{\Phi} \optilde{c}_p^\dagger.
\end{equation}
The one-body part in the objective function in \eqref{eq:oriopt-oriH}
then admits,
\begin{equation} \label{eq:equivalence-one-body}
    \begin{split}
        & \ev**{\sum_{p,q=1}^\morb h_{pq} \op{c}_p^\dagger
        \op{c}_q}{\Phi} \\
        = & \sum_{p,q=1}^\morb h_{pq} \ev**{\optilde{c}_p^\dagger
        \optilde{c}_q}{\Phi} \\
        = & \sum_{p,q=1}^\morb h_{pq} \sum_{p',q'=1}^\norb
        \ev**{\op{d}_{p'}^\dagger \op{d}_{q'}}{\Phi} U_{pp'} U_{qq'} \\
        = & \ev**{\sum_{p',q'=1}^\norb \tilde{h}_{p'q'}
        \op{d}_{p'}^\dagger \op{d}_{q'}}{\Phi}, \\
    \end{split}
\end{equation}
where $\tilde{h}_{p'q'}$ is defined as \eqref{eq:hpq}.  The one-body
part in the objective function in \eqref{eq:oriopt-oriH}, hence,
is equivalent to that in \eqref{eq:oriopt}.

In order to show the equivalence of the two-body part in both
objective functions, we need two more anticommutation relations.
The anticommutation relation between $\op{c}_s$ and $\optilde{c}_r$
satisfies,
\begin{equation} \label{eq:acommcctilde}
    \begin{split}
        \acomm{\op{c}_s}{\optilde{c}_r} = &
        \acomm{\op{c}_s}{\sum_{r'=1}^\norb \op{d}_{r'} U_{rr'}} \\
        = & \sum_{r'=1}^\norb \sum_{r'' = 1}^\morb \acomm{\op{c}_s}{
        \op{c}_{r''} } U_{rr'} U_{r''r'}\\
        = & 0.
    \end{split}
\end{equation}
Similarly, we also have the anticommutation relation between
$\optilde{c}_p^\dagger$ and $\op{c}_q^\dagger$,
\begin{equation} \label{eq:acommcdaggerctildedagger}
    \acomm{\optilde{c}_p^\dagger}{\op{c}_q^\dagger} = 0.
\end{equation}
The anti-commutation relations within $\optilde{c}$s can also be
derived in an analog way. The two-body part in the objective function
in \eqref{eq:oriopt-oriH} then admits,
\begin{equation} \label{eq:equivalence-two-body}
    \begin{split}
        & \ev**{\sum_{p,q,r,s=1}^\morb v_{pqrs} \op{c}_p^\dagger
        \op{c}_q^\dagger \op{c}_s \op{c}_r}{\Phi} \\
        = & \sum_{p,q,r,s=1}^\morb v_{pqrs} \ev**{\optilde{c}_p^\dagger
        \op{c}_q^\dagger \op{c}_s \optilde{c}_r}{\Phi} \\
        = & \sum_{p,q,r,s=1}^\morb v_{pqrs} \ev**{\op{c}_q^\dagger
        \optilde{c}_p^\dagger \optilde{c}_r \op{c}_s}{\Phi} \\
        = & \sum_{p,q,r,s=1}^\morb v_{pqrs} \ev**{\optilde{c}_p^\dagger
        \optilde{c}_q^\dagger \optilde{c}_s \optilde{c}_r}{\Phi} \\
        = & \sum_{p',q',r',s'=1}^\norb \sum_{p,q,r,s=1}^\morb v_{pqrs}
        U_{pp'} U_{qq'} U_{rr'} U_{ss'} \cdot \\
        & \cdot \ev**{\op{d}_{p'}^\dagger \op{d}_{q'}^\dagger \op{d}_{s'}
        \op{d}_{r'}}{\Phi} \\
        = & \ev**{\sum_{p',q',r',s'=1}^\morb \tilde{v}_{p'q'r's'}
        \op{d}_{p'}^\dagger \op{d}_{q'}^\dagger \op{d}_{s'}
        \op{d}_{r'}}{\Phi}, \\
    \end{split}
\end{equation}
where the second equality applies the anticommutation relations in
\eqref{eq:acommcctilde} and \eqref{eq:acommcdaggerctildedagger},
the third equality applies the anticommutation relations
within $\optilde{c}$s, and $\tilde{v}_{p'q'r's'}$ is defined as
\eqref{eq:vpqrs}.  The two-body part in the objective function
in \eqref{eq:oriopt-oriH}, hence, is equivalent to that in
\eqref{eq:oriopt}.

Combining \eqref{eq:equivalence-one-body} and
\eqref{eq:equivalence-two-body}, we conclude that the objective
functions in \eqref{eq:oriopt-oriH} and \eqref{eq:oriopt} are
equivalent given the wave function $\ket{\Phi} \in \Dphi$.

\newpage

\section{\ch{N2} Binding Curve}
\label{app:bindingcurve}

The \ch{N2} binding curve is plotted in Figure~\ref{fig:n2-bindingcurve} and
the detailed energies are given in Table~\ref{tab:n2-bindingcurve1} and
Table~\ref{tab:n2-bindingcurve2}. Table~\ref{tab:n2-bindingcurve1} provides
the ground-state energies for \ch{N2} with bond length smaller than that at
equilibrium geometry, whereas Table~\ref{tab:n2-bindingcurve2} provides the
ground-state energies with bond length greater than that at equilibrium
geometry. In both tables, we apply OptOrbFCI to compute the ground-state
energies of \ch{N2} under the cc-pVQZ basis set with 28 selected orbitals.
The same list of bond lengths as that in \citet{Wang2019} is adopted here.
The ground-state energies of FCI under the cc-pVDZ basis set are cited from
\citet{Wang2019}.

\begin{table}[h]
    \centering
    \begin{tabular}{ccc}
        \toprule
        Bond & FCI & OptOrb \\
        Length ($a_0$) & cc-pVDZ (Ha) & cc-pVQZ(28) (Ha) \\
        \toprule
        1.500 & $-108.6300476$ & $-108.8144031$ \\
        1.550 & $-108.7719968$ & $-108.9391824$ \\
        1.600 & $-108.8888460$ & $-109.0429050$ \\
        1.650 & $-108.9843136$ & $-109.1260152$ \\
        1.700 & $-109.0615754$ & $-109.1926550$ \\
        1.750 & $-109.1233484$ & $-109.2443696$ \\
        1.800 & $-109.1719641$ & $-109.2830113$ \\
        1.850 & $-109.2094264$ & $-109.3137005$ \\
        1.900 & $-109.2374578$ & $-109.3359754$ \\
        1.950 & $-109.2575411$ & $-109.3511562$ \\
        2.000 & $-109.2709530$ & $-109.3603603$ \\
        2.050 & $-109.2787896$ & $-109.3645818$ \\
        2.100 & $-109.2819938$ & $-109.3647561$ \\
        2.118 & $-109.2821727$ & $-109.3639435$ \\
        \bottomrule
    \end{tabular}
    \caption{Ground state energies for \ch{N2} with bond lengths
    smaller than $2.118 a_0$.} \label{tab:n2-bindingcurve1}
\end{table}

\begin{table}[h]
    \centering
    \begin{tabular}{ccc}
        \toprule
        Bond & FCI & OptOrb \\
        Length ($a_0$) & cc-pVDZ (Ha) & cc-pVQZ(28) (Ha) \\
        \toprule
        2.118 & $-109.2821727$ & $-109.3639435$ \\
        2.150 & $-109.2813737$ & $-109.3614955$ \\
        2.200 & $-109.2776211$ & $-109.3555534$ \\
        2.250 & $-109.2713283$ & $-109.3473397$ \\
        2.300 & $-109.2630013$ & $-109.3373910$ \\
        2.350 & $-109.2530718$ & $-109.3259793$ \\
        2.400 & $-109.2419079$ & $-109.3135325$ \\
        2.450 & $-109.2298228$ & $-109.2970156$ \\
        2.500 & $-109.2170830$ & $-109.2835331$ \\
        2.600 & $-109.1905077$ & $-109.2557994$ \\
        2.700 & $-109.1635998$ & $-109.2279395$ \\
        2.800 & $-109.1373583$ & $-109.2008712$ \\
        2.900 & $-109.1124729$ & $-109.1751472$ \\
        3.000 & $-109.0894053$ & $-109.1512795$ \\
        3.100 & $-109.0684502$ & $-109.1295149$ \\
        3.200 & $-109.0497787$ & $-109.1020069$ \\
        3.300 & $-109.0334619$ & $-109.0848174$ \\
        3.400 & $-109.0194835$ & $-109.0700669$ \\
        3.500 & $-109.0077466$ & $-109.0575569$ \\
        3.600 & $-108.9980829$ & $-109.0471986$ \\
        3.700 & $-108.9902691$ & $-109.0387653$ \\
        3.800 & $-108.9840499$ & $-109.0319008$ \\
        3.900 & $-108.9791625$ & $-109.0265483$ \\
        4.000 & $-108.9753572$ & $-109.0223481$ \\
        4.100 & $-108.9724102$ & $-109.0189859$ \\
        4.200 & $-108.9701316$ & $-109.0164481$ \\
        4.300 & $-108.9683664$ & $-109.0144948$ \\
        4.400 & $-108.9669909$ & $-109.0129196$ \\
        4.500 & $-108.9659102$ & $-109.0117220$ \\
        \bottomrule
    \end{tabular}
    \caption{Ground state energies for \ch{N2} with bond lengths
    larger than $2.118 a_0$.}
    \label{tab:n2-bindingcurve2}
\end{table}



\end{document}